# Metallo-Dielectric Photonic Crystals and Bandgap Blue-Shift

Alex Lonergan[1], Breda Murphy[1], and Colm O'Dwyer[1,2,3,4]*

*[1]School of Chemistry, University College Cork, Cork, T12 YN60, Ireland*

*[2]Micro-Nano Systems Centre, Tyndall National Institute, Lee Maltings, Cork, T12 R5CP, Ireland*

*[3]AMBER@CRANN, Trinity College Dublin, Dublin 2, Ireland*

*[4]Environmental Research Institute, University College Cork, Lee Road, Cork T23 XE10, Ireland*

## Abstract

One of the most appealing aspects of photonic crystal structures is the photonic bandgap created in structures with sufficiently high dielectric contrasts between constituent materials. Understanding how specific photonic crystal structures and their associated stopband positions can selectively interfere with incoming light is vital for implementing these structured dielectrics in a range of optical applications. Metallo-dielectric photonic crystals act to incorporate metal particles into the ordered arrangement of these structures. We examined copper, nickel and gold metal infiltration into polystyrene opals and TiO2 inverse opals. We report a consistent optical phenomena directly associated with the creation of metallo-dielectric photonic crystal structures. More pronounced and numerous diffraction resonances emerge in opal photonic crystals with a metal deposited across the top layer. Common to both opal and inverse opal structures, was a blue-shift in the position of the (111) photonic stopband which increased in magnitude with greater metal content in the structure. We investigate the origin of the photonic stopband blue-shift by variation of the metal content and the placement of metal in the photonic crystal structure. Metal introduced to structured dielectric media tunes the photonic stopband by altering the effective dielectric constant of the photonic crystal.

# Introduction

The ability to direct and control light propagating through a medium is an essential criterion when designing an optical material. Inherently, photonic crystal materials possess a structural control over certain frequency ranges of electromagnetic radiation incident on the ordered layers of the material. From the first suggestions of photonic crystal materials as alternating dielectric layers[1] [2], there has been significant interest in understanding and application of the optical control offered by photonic crystal structures[3] [4] [5] [6]. Of particular interest is the photonic bandgap established in these structures. Depending on the desired optical performance, photonic structures can be designed and modelled as 1D[7] [8], 2D[9] [10] or 3D[11] [12] [13] dielectric contrast materials. For 3D structures, yablonovite[14], artificial opals and inverse opals[15] are all examples of photonic crystal materials. Photonic bandgaps can be engineered in both wavelength position and width of the depletion zone through control over the structural size or dielectric contrast of the materials[16] [17]. A high dielectric contrast or difference between constituent materials of the photonic crystal medium is necessary for achieving a full, omnidirectional photonic bandgap[15] [18]. Materials with an inadequate dielectric contrast still exhibit a partial reduction of light intensity in the frequency range of the photonic bandgap. These incomplete photonic bandgaps are often referred to as pseudo-photonic bandgaps or photonic stopbands[19] [20] [21].

The threshold for achieving a complete photonic bandgap is relatively large considering the dielectric constants of available materials. Artificial opals are structures generally consisting of spherical particles stacked and embedded in air medium; typical examples include spheres of polystyrene (PS), poly-methyl methacrylate (PMMA) and silica[22] [23]. These low dielectric materials form a low dielectric contrast versus the air background, generally yielding a pseudo-photonic bandgap[19] [24]. Metal oxides with a high dielectric contrast versus an air background are popular choices for inverse opal materials[25] [26] [27] [28]. Critically, the magnitude of the dielectric contrast (or refractive index contrast) will determine the appearance and width of the photonic bandgap for a given arrangement of material. Simulations of photonic bandgaps predict the appearance of an omnidirectional bandgap in an inverse opal structure for a refractive index contrast of 2.8 or greater[29] [30], difficult to achieve with many metal oxide materials.

Even in the absence of an omnidirectional bandgap, photonic crystal structures have found a wide variety of applications, utilising both the easy to achieve structural order of materials and the presence of the stopband. Structurally ordering materials into a photonic crystal arrangement creates a porous material with higher levels of exposed surfaces compared to bulk arrangements. Higher surface areas are ideal for electrochemical[31] [32] [33] and catalytic[34] [35] [36] applications, where inverse opal structures have been applied for improved material performances. The sensitivity associated with the wavelength position of the photonic stopband has created numerous applications for both artificial opal and inverse opal materials. Shifts in the position of the photonic stopband can elucidate changes made to the material environment. Modifications to the periodicity or dielectric contrast of the system induce changes to the wavelength position of the photonic stopband; hence, photonic crystal materials are popular choices as optical sensors[37] [38]. A range of medical[39] [40] [41], biological[42] [43] and chemical[44] [45] [46] research has successfully utilised the optical stopband of photonic crystals as sensors for detecting a variety of different substances. The optics of the ordered structures are not limited to just detection; the presence of an optical stopband across controllable wavelength regimes is ideal for creating optical waveguides[47] [48]. Light confinement by the dielectric contrast of the material structure has been used to develop optical cavities which have been applied as optical gain regions in lasing materials[9] [49]. Other spectral features associated with these materials, such as the slow photon effect[50] [51] at the edge of the optical bandgap region, can also be exploited to improve material performance in photocatalytic applications[52] [53].

There have been numerous studies characterising the properties of ordered material structures and their effects on the signature optical response of the photonic stopband[54] [55] [56] [57]. Choice of material dielectric constant for both high and low-index regions[55] [58] [59], angle of incidence[60] [61] [62], periodicity[60] [61] [62], structural anisotropy[63] [64] and liquid infiltration into the voids[56] [61] [65] are just some of the parameters which influence the optical behaviour of the structure. An interesting subclass of photonic materials are metallo-dielectric photonic crystals, where a modification of the base photonic crystal is accomplished through incorporation of a metal into the structure. It was initially proposed that the addition of metals into the photonic crystal structure should broaden the photonic bandgap due to the increase in the dielectric contrast of the composite material[66] [67]. At energies below the plasma frequency of a particular metal, the real part of the dielectric constant of that metal is negative; replacing some of the air in photonic crystal with a metal was expected to broaden the dielectric contrast between materials in the structure. Considering the geometry of the photonic crystal material, there are numerous methods of hosting metals in the ordered structure; material can be deposited on a metal layer, the top surface of the photonic crystal can be coated and a complete or partial infilling of the structural voids with metal are all possibilities[68]. Metallo-dielectric structures have been predicted to introduce some interesting optical behaviour to the photonic band structures of these materials[69] [70] [71]. For a metal film deposited on an opal template, it has been suggested that an optical cavity can form; higher amounts of reflections from the opal-metal interface increase the optical path length and can accentuate the appearance of diffraction resonances from lower index planes (non [111] planes) on the optical spectrum compared to an uncoated material[68]. Interestingly, there have been some reports of a blue-shift in the wavelength position of the (111) diffraction stopband arising from metal insertion into the opal structure[67] [68] [72] [73] [74] [75]. The general assumption of the origin behind this blue-shift in the photonic stopband is a reduction in the effective dielectric constant or refractive index of the photonic crystal structure[67] [72] [74] [75].

Here, we present an investigation into the structural and optical evolution of Cu, Au and Ni metal-coated photonic crystal structures. Metals are deposited directly on the surface of PS colloidal opals and $TiO_2$ inverse opal structures via physical vapour deposition. Microscopy analysis of the metallised surfaces are used to characterise the appearance and deposition pattern of the metal films. Several thicknesses of metal coatings are deposited on photonic crystals surfaces and the consistency and growth of the metal film on surface features are assessed. Changes to the optical transmission spectrum upon addition of a metal film is tracked for several film thicknesses and different metals. Principally, a blue-shift in the (111) diffraction stopband is detected and tracked for each metal deposited on the surface of the PhC. The magnitude of the blue-shift is monitored relative to the thickness of the metal film on the surface, with more pronounced wavelength shifts reported for thicker films. Both PS opals and $TiO_2$ inverse opals are shown to demonstrate a blue-shift in the main diffraction stopband upon addition of a metal. These observations support previous assertions[67] [72] [74] [75] that metals in photonic crystal structures act to enhance the dielectric contrast between materials and induce a reduction in the effective dielectric constant of the material. Crucially, the dielectric contrast of the metal material infilling the structure is shown to be necessary for the blue-shift in the case of Cu, where partial oxidation of the metal to a metal oxide is demonstrated to decrease the magnitude of the blue-shift. Our results show that even a moderate modification to the surface of an opal or inverse opal structure with a metal is sufficient to induce a non-trivial change in the signature optical response of the photonic crystal material. We attempt to quantify the magnitude of the spectral blue-shift with film thickness and relate the impact of the negative dielectric metal on the average refractive index of the overall structure.

## Experimental Methods and Materials

## Pre-treatment and Preparation of Substrates

A transparent and conductive substrate was used for template formation, ideal for both optical measurements and microscopy examination. Fluorine-doped tin oxide (FTO) coated soda-lime glass substrates were purchased from Solaronix SA and were used for in the formation of all samples in this work. The breakdown of the conductive coating can be described by a thin layer (~20 nm) of FTO, deposited on a thin layer (~20 nm) of $SiO_2$, all deposited on a final thicker layer (~300 nm) of FTO on a glass substrate. The substrate glass was 2.2 mm thick and larger pieces were cut to make sample surface areas of $10 \times 25$ mm$^2$. Glass substrates were cleaned via successive sonication for 10 minutes in acetone (reagent grade; 99.5% Sigma Aldrich), isopropyl alcohol (reagent grade; 99.5% Sigma Aldrich) and de-ionised water. Cleaned sample surfaces were allowed to dry at room temperature. On the conductive side of the glass substrates, sample areas of $10 \times 10$ mm$^2$ were formed using Kapton tape to cover the remainder of the glass surface; both the front and back of the glass surface was covered with Kapton tape to promote the formation of colloidal crystals on the conductive $10 \times 10$ mm$^2$ sample area only. A Novascan PSD Pro Series digital UV-ozone system was used to pre-treat glass substrates and improve the hydrophilicity of the surface for colloidal crystal template formation. Immediately prior to dip-coating procedure, prepared and cleaned sample surfaces were pre-treated by UV-ozone cleaning for 1 h.

## Polystyrene Opal Template Formation Procedure

Monodisperse PS spheres with a nominal diameter of approximately 370 nm were purchased in an aqueous suspension from Polysciences Inc. at a concentration of 2.5 wt% PS. Purchased sphere suspensions contained a slight negative charge owing to the sulfate ester used in the formation of the colloidal particles. Sphere suspensions for dip-coating template formation were used as received; sphere suspensions were maintained at the 2.5 wt% concentration. Prior to dip-coating, approximately 2 mL of sphere suspension was transferred to a small glass vial and pre-heated to a temperature of approximately 50 °C. UV-ozone pre-treated sample surfaces were dip-coated into preheated sphere suspensions at a rate of 1 mm min[-1] while the temperature of the suspension was held constant. An MTI PTL-MM01 dip coater was used to control the speed of insertion and withdrawal. Glass substrates were inserted at a slight incline (~ 10 – 20°) to the vertical in an attempt to improve adhesion of the PS colloids to the desired surface[76]. When the desired surface area of the substrate was full immersed in the suspension, the substrate was held still for 10 minutes to allow the sphere suspension to settle to a minimum energy state from surfactant-mediated repulsion. The colloidal crystal template was formed upon withdrawal from the suspension at a rate of 1 mm min[-1] and samples were allowed to dry in an inert atmosphere.

## TiO$_2$ Inverse Opal Formation

TiO$_2$ inverse opals were fabricated using PS opal samples as templates using the well-known sol-gel synthesis technique[27] [33] [62] [65] [77]. Opal sphere templates were infiltrated with liquid (sol) precursor and the air voids in the opal photonic crystal were filled with the precursor. Ambient moisture reacts with the precursor in the opal structure and material (gel) forms in the interstitial voids which can be crystallised at high temperatures. For TiO$_2$ inverse opals, a 0.1 M TiCl$_4$ solution in isopropyl alcohol was used as the precursor. An adequate amount of the precursor, enough to sufficiently wet the entire surface, was drop-cast onto the PS opal templates. The PS opal templates in the structure were removed via calcination in air at 450 °C with a ramp rate of 5 °C min[-1]. The high temperature also allowed for the crystallisation of the TiO$_2$ material to crystalline anatase phase TiO$_2$. A titanium (IV) chloride tetrahydrofuran complex (TiCl$_4$ · 2THF 97 %; Sigma Aldrich) was used as the source of TiCl$_4$.

## Formation of Metal-Coated Photonic Crystals

Metal films on the surfaces of photonic crystals were formed via physical vapour deposition, specifically, magnetron sputter coating with a Quorum 150T S magnetron sputtering system. Removable metal disc targets of thickness 0.1 mm and diameter 57 mm were used for copper, gold and nickel deposition. All targets were purchased from Ted Pella Inc. and featured a material purity of 99.99% for each metal. A film thickness monitor was used to deposit thicker metal coats onto surfaces; microscopy measurements were used to determine the actual thickness of the metal deposit formed due to the curved surfaces of photonic crystals. All metals were sputtered under an inert gas of high purity to ensure that the gas molecules did not react with the sputtered metal particles. Argon gas ($>$ 99.995% purity) was used as the sputtering gas in all cases. A target shutter in the sputter coater assembly allowed for cleaning of the target surface prior to film deposition, particularly important in the case of oxidising metal targets.

## Optical Transmission Analysis

Optical transmission spectra were recorded using an Ocean Optics Inc. UV-visible spectrometer (USB2000+ VIS-NIR-ES) with an operational range of 350 – 1000 nm. A tungsten-halogen lamp purchased from Thorlabs Inc. operating over the wavelength range 400 – 2200 nm was used as the unpolarised excitation source. The angle of incidence between the normal of the photonic crystal surface and the incident light was fixed at 0° for all measurements. For the transmission spectra measured for each material, 100% transmittance was normalised to the transmission of a clean blank piece of FTO. Unless otherwise stated, the transmission spectra of all metal coated samples were recorded immediately after the coating procedure to minimise oxidation effects. To minimise spectral variance across samples, particularly in the case of pre and post-coated metal surfaces, spectra shown for a particular sample depict the features present (namely the position of the photonic stopband) across an average of 10 separate transmission spectra taken in close proximity on the sample surface.

## Microscopy and Materials Characterisation

Scanning electron microscopy (SEM) was carried out using a Zeiss Supra 40 high resolution SEM at typical accelerating voltages of 15 – 20 kV. The instrument was used to analyse the morphology of the opals and IO PhCs, and the metal film morphology or thickness. Dimensional analysis for feature size distributions from SEM images were measured using IMAGEJ software. Energy dispersive X-ray spectroscopy (EDX) analysis was also performed on the same SEM instrument using an accelerating voltage of 15 kV. EDX line scan measurements were obtained in several directions across the surface and used to assess the consistency of the metal coating on the photonic crystal structure.

## Results and Discussion

## Metallo-Dielectric Copper-Coated Polystyrene Opal Photonic Crystals

To assess the optical response associated with these metallo-dielectric photonic crystals, we first characterised the Cu shell formed on the surface of the PS opal photonic crystals. The thickness of the Cu film was controlled by varying deposition time during sputtering and correlating SEM thickness and metal film morphology with the values from the quartz crystal monitor.

Figure 1 (a) shows a schematic diagram that depicts the shape of the metal layer formed on the top surface of the opal. From a top-down 2D perspective, as would be observed on a conventional SEM image, the circular top shape of the PS sphere is gradually converted to a hexagonal shell with increasing metal content. The hexagonal shaped metal shell should fit like a cap on top of the spherical opal. For an ordered FCC packed opal structure, the top layer should feature spherical PS spheres in point contact with one another as expected from the (111) planar geometry. As the mount of deposited metal increases during sputtering, from a 2D perspective, the hexagonally shaped metal layer expands into the interstitial air voids of the metallo-dielectric photonic crystal structure. In this scenario, point contacts between neighbouring spheres are converted to linear contact regions between neighbouring metal shells. With the metal layer overlaid on top of the initial opal template, as in Fig. 1 (a), images of the metal-coated structure can be used estimate both the initial opal diameter ($D_{sphere}$, measured from opposing hexagonal sides) and the diameter of the metal coating on the surface ($D_{shell}$, measured from opposing hexagonal vertices).

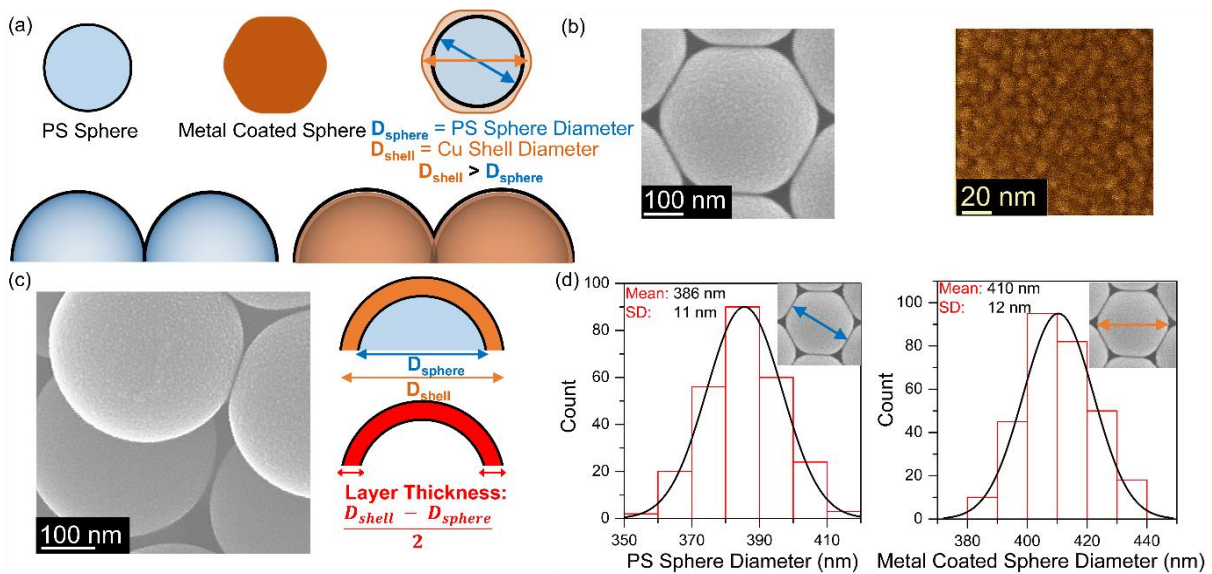

**Fig. 1** (a) Plan-view schematic of the hexagonal PS sphere transformation after metallisation of the top surface of the colloidal opal and (b) SEM image of coated PS spheres. (c) Schematic representation of the definition used for the metal layer thickness with an SEM image showing the preferential coating of the top layer of opals. The thickness of the metal layer is obtained from SEM analysis as half of the difference between the metal shell diameter ($D_{shell}$) and the initial opal diameter ($D_{sphere}$). (d) Size distributions for metal-coated PS spheres showing the average initial sphere diameter before and after metal coating.

Figure 1 (b) displays SEM images depicting the geometric changes to the top surface layer of the opal when a Cu metal layer is deposited. The SEM images confirm the appearance of the metal-modified structure as posited from the schematic in Fig. 1 (a). The spherical shape of the PS sphere is converted to a hexagonal shape when coated even with a moderate thickness metal layer. A metal layer of just 12 nm thickness atop a 386 nm PS opal is seen to sufficiently alter the geometry of the all the spheres in the top layer of the opal. The metal coating on the surface of the structure is found primarily on the top hemisphere of each sphere. The SEM image in Fig. 1 (c) displays a side profile of Cu metal coated onto a PS opal. Even from a side profile, the preferential coating of the top surface with metal is observed. Opal layers beneath the top surface do not feature a complete conformal metal coating of Cu.

Figure 1 (d) illustrates the method used to calculate the average metal layer thickness deposited across the opal structure. Diameter size distributions were constructed from the SEM images of the metallo-dielectric structures. Several hundred diameter measurements were made across multiple SEM images of the sample surface for both $D_{sphere}$ and $D_{shell}$. Thus type of approach allows a mean value for each diameter to be calculated and also assesses any potential asymmetry in the metal coatings. A mean diameter is useful for optical analysis where the periodicity of the structure is vital for predicting the optical response of the photonic stopband. More details on the measurements of $D_{sphere}$ and $D_{shell}$ for Cu metal coatings of various thicknesses can be found in the Supplementary Materials Fig. S1. From these size distributions, Cu metal layers of 2.5, 8, 12 and 15 nm were calculated to be deposited on opal surfaces. The appearance of the Cu metal layer on the surface of the opal with various different layer thicknesses can also be found in the Supplementary Materials Fig. S2. With increasing metal film thickness, there is a visible reduction in the interstitial area occupied by air in the structure.

As the metal layer expands on the opal surface the interstitial area is reduced accordingly, and determined to be proportional to the thickness of the metal coating. The Supplementary Materials Fig. S3 characterises this reduction in interstitial area, as observed from SEM images, for each thickness of the Cu metal layers deposited on an opal surface.

Analysis of the SEM images of metal coated opals is useful for providing a quantitative basis for assessing the amount of metal deposited on the surface as the metal layer thickness increases. Next, the optical effects of Cu metal deposited on the surface of an opal photonic crystal is assessed. The spectral response of the Cu metal film atop the opal surface can be interpreted relative to the uncoated opal in terms of a reduction in the overall transmission of the modified photonic crystal structure. The reduction in transmission, the appearance of further diffraction resonances and the blue-shift of the principal optical stopband become more prominent with increasing metal layer thickness. These effects associated with increasing metal content in the photonic crystal structure are presented in Fig. 2.

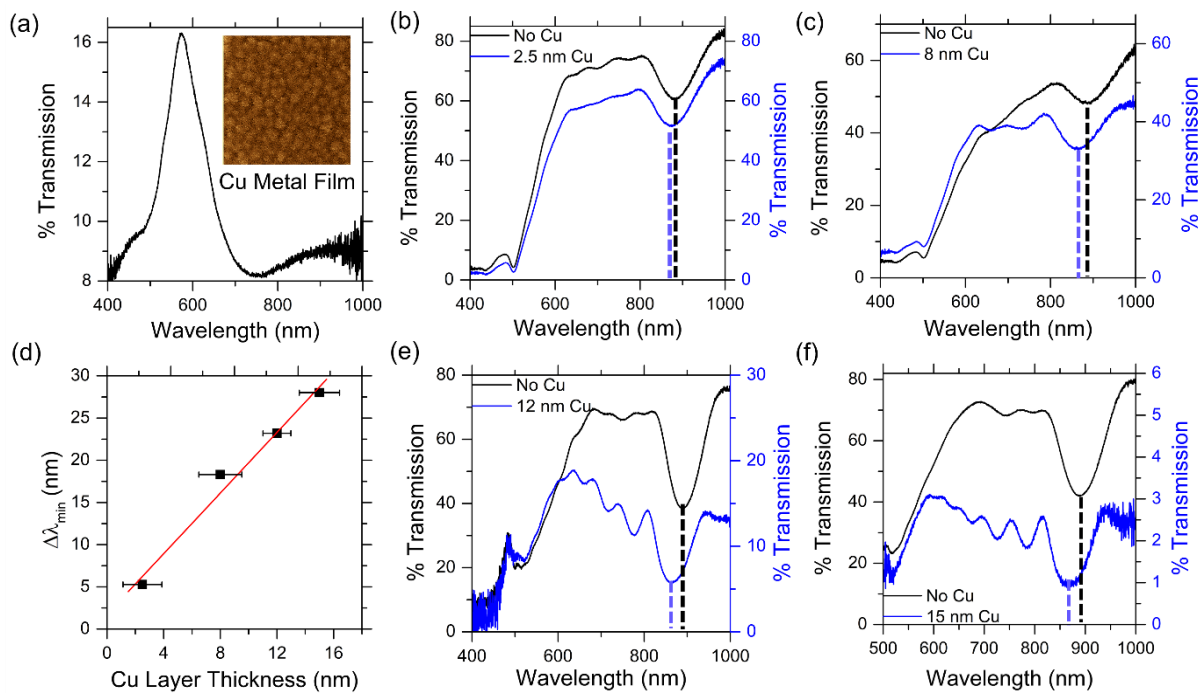

**Fig. 2** (a) A transmission spectrum for a flat Cu film deposited on a glass substrate using the same coating speeds and deposition times used for a 15 nm Cu metal layer on an opal surface. All optical transmission spectra were recorded within 5 minutes of the sputter coating process in order to minimise oxidation of the Cu metal. Transmission spectra for bare (black lines) and Cu-coated (blue lines) opals shown for metal layer thicknesses of (b) 2.5 nm, (c) 8 nm, (e) 12 nm and (f) 15 nm. Note the significant reduction in transmission intensity for coated metallo-dielectric PhCs. (d) Measured photonic stopband blue-shift ($\Delta\lambda_{min}$) as a function of Cu layer thickness.

Figure 2 (a) displays the transmission spectrum associated with a flat Cu metal film on FTO-coated glass deposited under the same conditions used to form a 15 nm Cu metal layer on an opal surface. The substantial attenuation of the light intensity, seen as a significant drop in the transmission intensity, is as expected for a highly reflective metal film on a surface. For thin films of Cu metal, the transmission maximum at approximately 570 nm and slight transmission minimum at approximately 760 nm have been previously attributed to interband photon transitions for fully filled $d$-shells and a localised plasmon effect depending on the thickness of the film, respectively[78]. The high reflectivity of the metal coating results in the decreasing transmittance of metal coated opal structures with increasing metal layer thicknesses. This trend can be observed in Figs. 2 (b), (c), (e) and (f) where the light intensity observed in transmission is strongly attenuated with increasing Cu layer thickness. From these optical transmission spectra, the appearance of non [111] diffraction resonances from the opal structure become much more pronounced with the addition of a Cu metal layer over the top surface of the opal. This is particularly apparent in Figs. 2 (e) and (f) where the thickest Cu metal layers are deposited. For thicker metal layers, 4 diffraction resonances become prominent transmission dips in comparison with the uncoated opal sample. It has been previously proposed that the emergence of non [111] diffraction resonances in metal-coated opal structures can be attributed to the formation of an optical cavity, enhancing the relative intensity of higher energy resonances compared to the overall transmission of the photonic crystal[68]. This explanation seems plausible in our case, where significant spectral features emerge at the high energy side of the principal (111) diffraction stopband.

The wavelength position for the transmission minimum ($\lambda_{hkl}$) associated with diffraction from a crystal plane of indices hkl can be calculated from the Bragg-Snell relation as follows:

$$\lambda_{hkl} = \frac{2d_{hkl}}{m} \sqrt{n_{eff}^2 - n_{sol}^2 \sin^2\theta}$$ (1)

Referring to Eq. (1), $d_{hkl}$ is the associated interplanar spacing of the crystal plane, $m$ is the diffraction resonance order, $n_{eff}$ is the effective refractive index of the photonic crystal structure and $\theta$ is the angle of incidence between the normal to the crystal plane and the incident light. For the spectra presented in Fig. 2, some simplifications to Eq. (1) can be made by assuming a first order resonance ($m = 1$) and normal incidence ($\theta = 0°$) of the incoming light. For these conditions the Bragg-Snell relation for the (111) diffraction plane becomes:

$$\lambda_{111} = 2\, d_{111}\, n_{eff}$$ (2)

For an FCC structure, as anticipated from an artificial opal template, the interplanar spacing ($d_{111}$) of the (111) plane can be determined using the centre-to-centre (D) distance between opals in the structure are follows:

$$d_{111} = \sqrt{\frac{2}{3}}\, D$$ (3)

The effective refractive index for a composite structure, such as a photonic crystal medium, can be estimated using a number of different methods[65]. For an artificial opal template consisting of materials with high ($n_1$) and low ($n_2$) refractive indices in respective volume fractions $\varphi_1$ and $\varphi_2$, the effective refractive index can be calculated as:

$$n_{eff} = n_1\varphi_1 + n_2\varphi_2$$ (4)

Thus, the wavelength position of the principal (111) diffraction stopband in an uncoated FCC PS ($n \sim 1.57$ at 900 nm[79], $\varphi = 0.74$) opal in an air background ($n = 1$, $\varphi = 0.26$) can be determined from knowledge of the centre-to-centre diameter as follows:

$$\lambda_{111} = 2.84 \sqrt{\frac{2}{3}} D \qquad\qquad (5)$$

The mean centre-to-centre distances, as calculated from the size distributions for each sample, are adopted here as an estimate of the average D for each opal sample. For the uncoated opal samples presented in Figs. 2 (b), (c), (e) and (f) the mean centre distances were calculated as 394, 386, 386 and 393 nm, respectively. Estimating the position of the principal (111) transmission stopband using Eq. (5) yields wavelength position estimates of 913, 895, 895 and 911 nm, respectively. These estimates show a relatively strong agreement with the experimentally determined stopband positions of 875, 875, 885 and 888 nm, shown in Figs. 2 (b), (c), (e) and (f), respectively. The metallo-dielectric opals in these spectra all display a blue-shift in the position of the photonic stopband, and the magnitude of the blue-shift increases for thicker Cu layer deposits. In the case of the Cu-coated opal spectra, shown in blue in Figs. 2 (b), (c), (e) and (f), the central wavelengths of the shifted photonic stopbands are now located at 870, 857, 862 and 860 nm, respectively. The application of a conformal Cu metal coating to the top layer of the opal surface with layer thicknesses of 2.5, 8, 12 and 15 nm induced a blue-shift in the photonic stopband of 5, 18, 23 and 28 nm, respectively. Figure 2 (c) displays this relation graphically, where a thicker Cu metal layer is shown to induce a greater shift in the photonic stopband, in general. In the proceeding sections the existence of this blue-shift is investigated for other metals and photonic crystal structures with a mechanism proposed for its origin.

**Metallo-Dielectric Cu, Au and Ni-Coated Opal and Inverse Opal Structures**

By variation of the metal coating material and the type of photonic crystal surface, the optical effects of metal infiltration into a dielectric medium are further investigated, specifically focusing on any changes to the position of the principal (111) photonic stopband. The optical effects of metal coatings of Cu, Au and Ni on both PS opal templates and $TiO_2$ inverse opals are assessed here. First starting with the coated PS opal structures, Figs. 3 (a), (b) and (c) display the optical transmission spectra for Cu, Ni and Au metal coatings deposited on the top surface of the opal, alongside SEM images of the coated structure. The thicknesses of the metal layers deposited were determined from SEM analysis using the same process described in Fig. 1. More detailed analysis of the SEM images and size distribution data for the Ni and Au coated PS opals can be found in the Supplementary Materials Figs. S4 and S5. Metal layers of Cu, Ni and Au were found to be deposited at thicknesses of 12, 9 and 5.5 nm. For all metals, a complete surface coating of the opal is formed and from the SEM images the metal shell creates the 2-D hexagonal pattern as discussed for the Cu metal. Figures 3 (a), (b) and (c) also show higher magnification SEM images for the Cu, Ni and Au metal layers. Similar sized metal nanoparticles are found in the Cu and Ni films (~ 5 – 8 nm) with slightly larger metal nanoparticles found for the Au film (~10 – 15 nm).

The optical transmission spectra for uncoated PS opals and metal-coated opals are also shown for Cu, Ni and Au layers in Figs. 3 (a), (b) and (c), with the spectral appearance of a flat metal film shown underneath the plot. The high reflectance of the metal films in the visible spectrum can be seen from the flat films deposited on the glass surface, with low transmission intensity through the films. The Au metal film displays the highest transmission intensity, presumably due to it being the thinnest deposit of metal on the surface. In the case of the 12 nm thick Cu-coated opal, the position of the photonic stopband has already been discussed in the previous section. The uncoated opal used in the deposition of Cu metal displays a photonic stopband at 885 nm, close to the predicted position of 895 nm from Eq. (5) using a mean diameter of 386 nm. The mean diameters for the uncoated opals in the case of the Ni and Au deposition, as calculated from SEM analysis and size distributions in Supplementary Materials Fig. S5, were calculated as 376 and 372 nm respectively. The predicted wavelengths of the photonic stopbands for Ni and Au uncoated opals, as per Eq. (5), are calculated as 872 and 863 nm. From the transmission spectra in Figs. 3 (b) and (c), there is a reasonable agreement between these calculated positions and the experimentally observed stopbands at 896 and 897 nm for the uncoated opals used for Ni and Au deposition.

With a Cu layer of 12 nm deposited on the opal surface, we measured a blue-shift of the principal (111) diffraction stopband from 885 nm to 862 nm after metallization. Similarly, this stopband blue-shifting is observed with the addition of both Ni and Au metal. For a nominal Ni metal layer thickness of 9 nm, the photonic stopband shifts from 896 nm to 878 nm. Even a thin 5.5 nm metal layer of Au induces a moderate stopband blue-shift of 15 nm, from 897 to 882 nm. In all instances, metal deposited on the opal surface is demonstrated to induce a blue-shift of the (111) diffraction stopband with a significant reduction in transmission intensity and the emergence of higher energy cavity resonances.

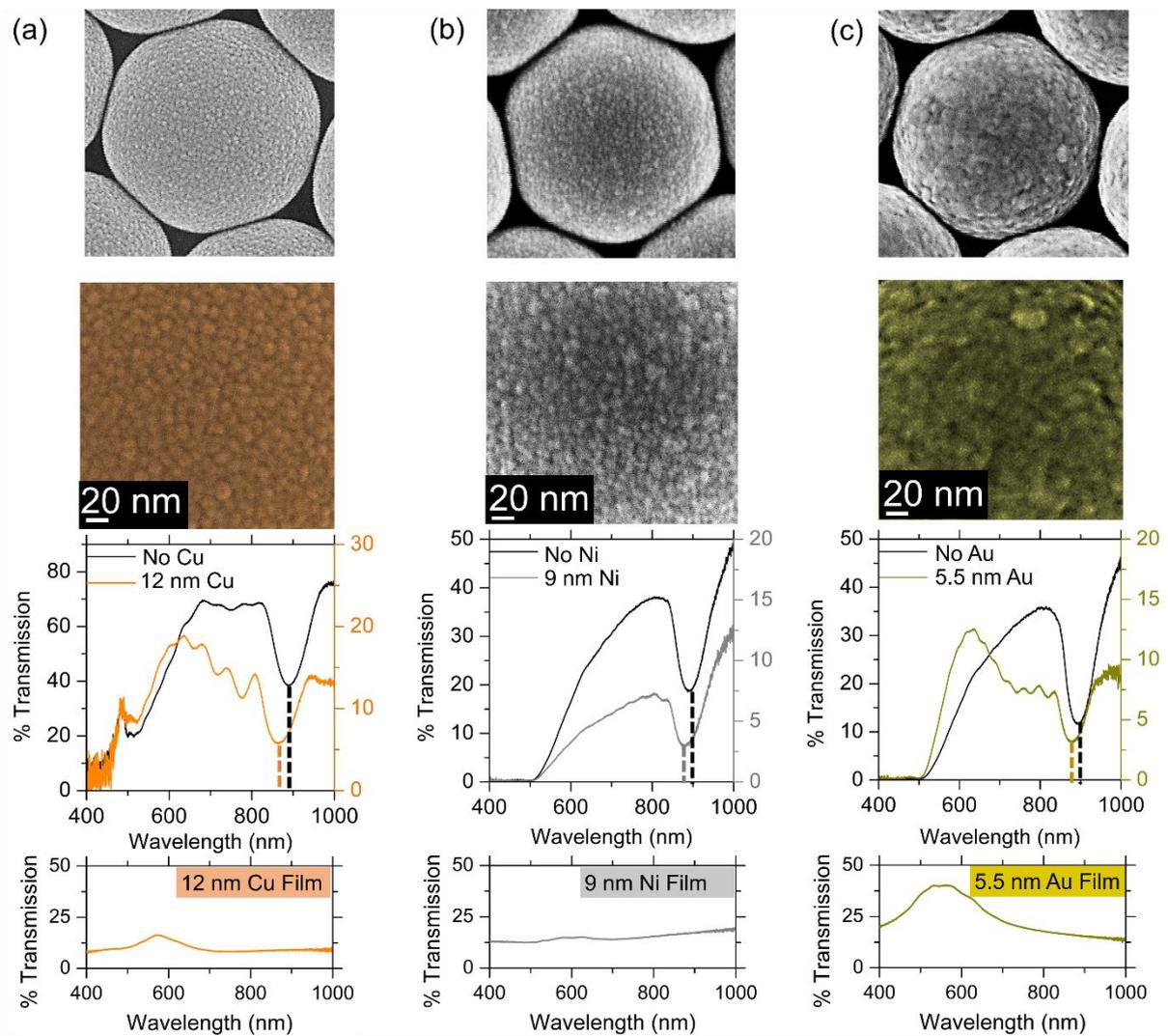

**Fig. 3** Series of data showing SEM images at low and high magnifications for metal films on opal surfaces, the optical spectra for uncoated opals compared to metal-coated structures and the transmission spectra associated with a flat metal film deposited on a glass substrate for (a) a 12 nm Cu layer, (b) a 9 nm Ni layer and (c) a 5.5 nm Au layer deposited on the top surface of an opal.

We also examined the blue-shift phenomenon with metallo-dielectric photonic crystal fabrication with the inverse opal geometry using Cu, Ni and Au metals. The deposition rates and times used in the formation of the metal coats were identical to the parameters used to coat opal surfaces in Fig. 3. The structure of the IO is inherently different to the parent opal template; TiO$_2$ IOs consist of low dielectric air spheres ($\varphi = 0.74$) surrounded by a network of high dielectric TiO$_2$ material ($\varphi = 0.26$), with a specific overall effective refractive index[62] [65].

With a metal deposited onto an IO material, it stands to reason that the metal can penetrate deeper into the material compared to an opal structure due to the highly porous structural geometry of the IO.

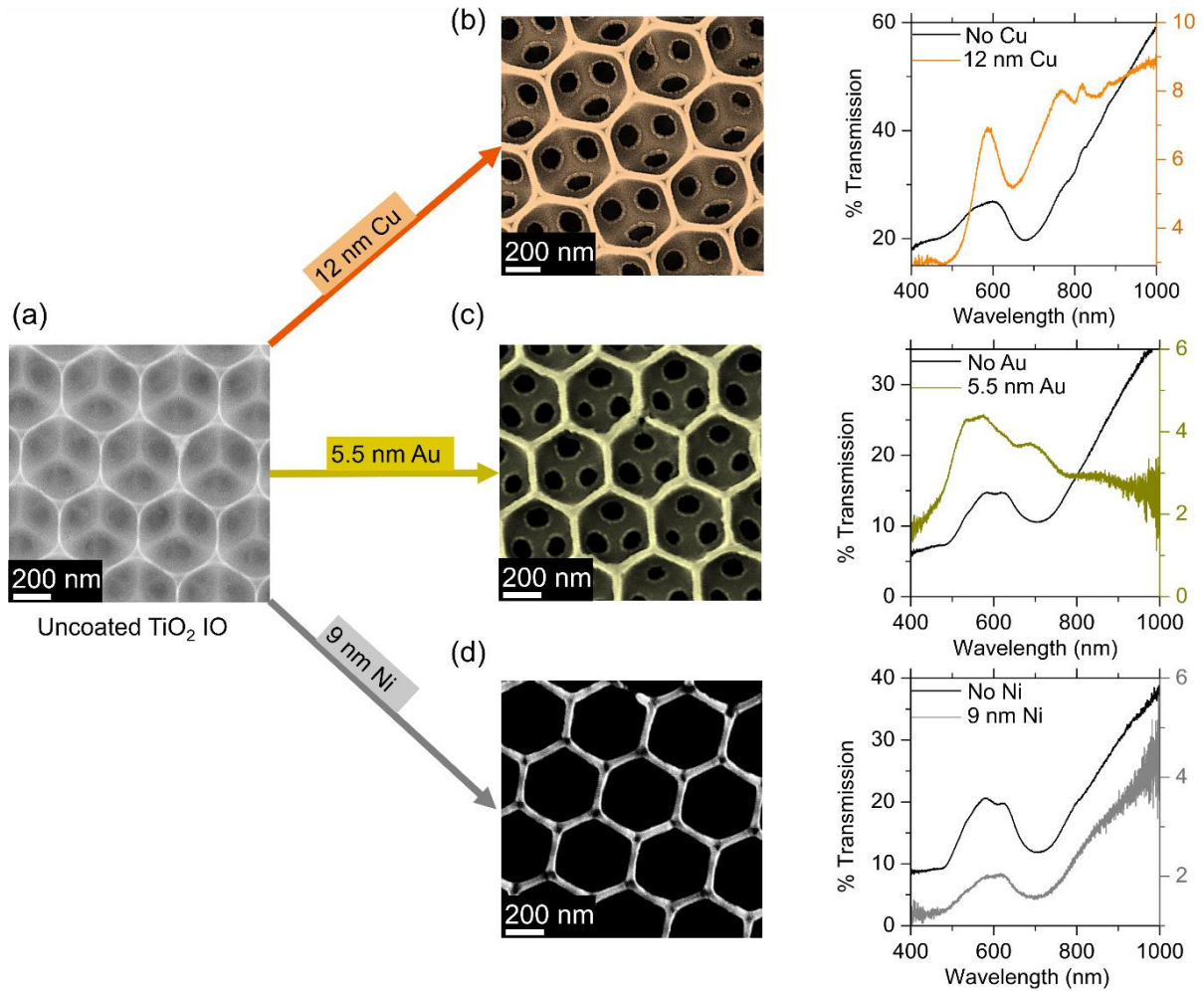

**Fig. 4** (a) The pristine surface typical to SEM images of uncoated TiO$_2$ inverse opal prepared from sol-gel infiltration of a PS opal template. SEM images of metal-coated TiO$_2$ IOs shown alongside a comparison of the optical spectra between uncoated TiO$_2$ IOs and metal-coated TiO$_2$ IOs for (b) Cu, (c) Au and (d) Ni metal sputtered onto the surface of the IO. The thicknesses of the metal layers are the same deposition parameters for the coating.

The effects of Cu, Ni and Au metal deposition onto a TiO$_2$ IO surface can be seen in Fig. 4. The surface of a typical TiO$_2$ IO can be seen in the SEM image in Fig. 4 (a), showing the ordered macroporous IO structure formed from inverting the planar (111) PS opal crystal plane. The mean pore size associated with TiO$_2$ IOs, as measured from size distribution data shown in the Supplementary Materials Fig. S6, is calculated to be 440 nm. Previous studies[62] [65] of the optical characteristics of $TiO_2$ IOs of similar dimensions have indicated the presence of an optical stopband located at ~700 nm. The transmission spectra for the uncoated $TiO_2$ IOs shown in Figs. 4 (b), (c) and (d) feature optical stopbands located at 687, 709 and 702 nm, respectively; the position of the photonic stopband is relatively consistent for similarly sized material pores.

The SEM images in Figs. 4 (b), (c) and (d) for $TiO_2$ IOs coated with Cu, Au and Ni show the effects of the metal deposition on the IO structure. The thin walls (~12 nm) of $TiO_2$ material on the top surface become thicker (~45, 32 and 35 nm for Cu, Au and Ni, respectively) as metal deposits over the structure. The deposited metal can also be seen to coat beneath the top surface with metal particles clearly visible on layers below the surface. The SEM images support the assumption that the sputtered metal particles can penetrate further into the porous IO material and incorporate into the structure beyond the top surface. Additional SEM images of metal coated $TiO_2$ IOs can be found in the Supplementary Materials Fig. S7 for further characterisation of the modified structure.

The positions of the photonic stopbands for Cu, Au and Ni metallo-dielectric $TiO_2$ IOs can also be seen from the transmission spectra presented in Figs. 4 (b), (c) and (d). Similar to the metal-coated opal structures, the position of the photonic stopband is blue-shifted to some extent for every $TiO_2$ IO modified with a metal deposit. In the case of the Cu-coated $TiO_2$ IO, there is significant shift in the stopband position to higher energy (from 687 nm to 651 nm, a blue-shift of 36 nm) when metallised with Cu. The photonic stopband is clearly visible for the Cu-coated $TiO_2$ IO. In contrast, the photonic stopband for the Au-coated $TiO_2$ IO is more difficult to observe on the transmission spectrum, due to partial overlap of the broad transmission maximum observed for Au films with the shifted stopband in this case. The small transmission dip observed at 662 nm is most probably the location of the shifted stopband for the Au-coated $TiO_2$ IO, constituting a significant blue-shift of 40 nm. The Ni-coated $TiO_2$ IO exhibits the smallest shift in photonic stopband position, shifting from 709 nm to just 701 nm, a blue-shift of 8 nm. The magnitude of the larger blue-shifts observed in both Cu and Au could be explained by the plasmonic properties of these metals and the injection of hot electrons from $Cu^{[80]}$ and $Au^{[81]}$ into $TiO_2$; hot electron effects are known to enhance the photocatalytic properties of semiconductors such as $TiO_2$. Nevertheless, in all cases of metal incorporation into IO structures, a blue-shift of the photonic stopband is observed. This blue-shift persists in both opals and IOs in spite of the inverse porous structure and energies of their characteristic stopbands. We next attempt to account for this shifting of the photonic stopband by analysing the influence of metal type and its placement in the structure.

## The Significance of Metal Particles for the Photonic Stopband Blue-Shift

Having observed a blue-shift in the photonic stopband position for all photonic crystal structures coated with a metal layer, the significance of the metal inclusion in the structure is next explored. Cu metal is known to oxidise slowly at room temperature to form copper oxides; the oxidation process is continuous with dominant $Cu_2O$ and minor $CuO$ phases of oxidised copper formed[82]. In the case of metallo-dielectric photonic crystals, this oxidation process can be leveraged to investigate how the metal specifically alters the optical properties of the structure. The oxidation of the metal over time effectively removes a certain percentage of metal material as it is replaced with an oxide of the metal. The position of the photonic stopband for photonic crystal structures is susceptible to slight variations between similarly sized structures; this can be observed even examining the data presented in Fig. 1, where the positions of the principal photonic stopband fluctuated slightly for similarly sized opal templates. An oxidation of Cu metal on the surface of an opal template allows for a variable metal content to be examined on the exact same template, removing any possibility of potential stopband position fluctuations between different samples.

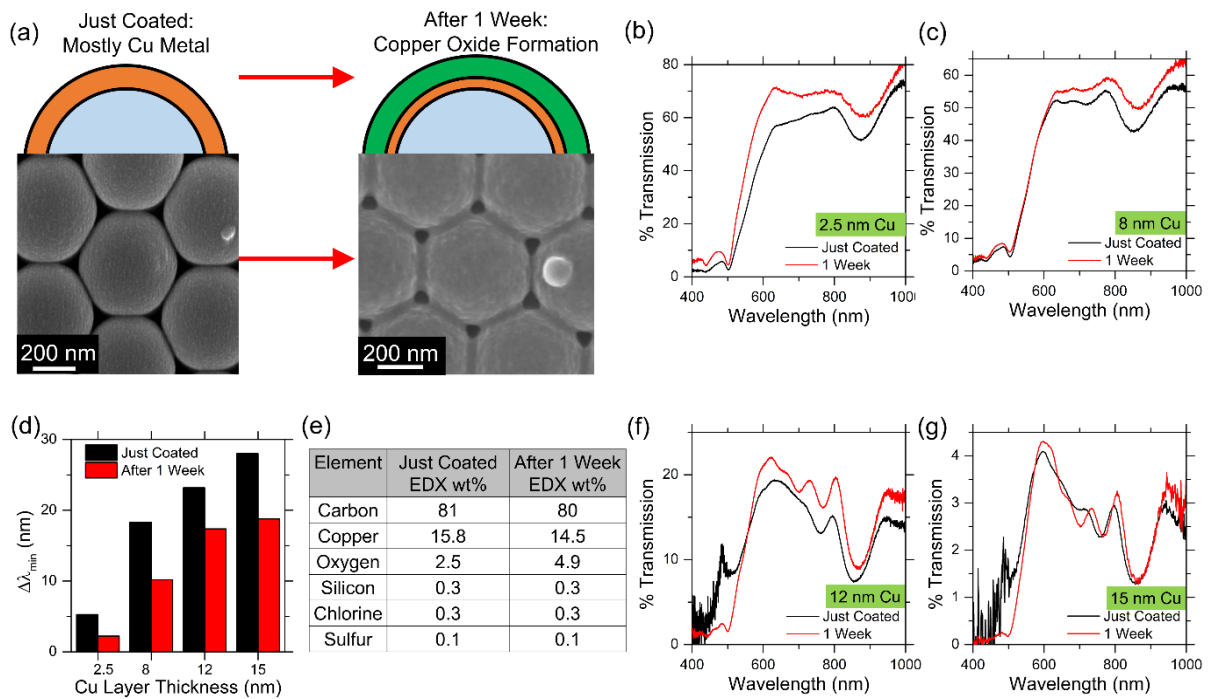

**Fig. 5** (a) A schematic representation of copper oxide formation on the surface of Cu metal films over time. Transmission spectra showing a comparison between newly coated Cu metal layers and the same samples after 1 week in ambient atmosphere for (b) 2.5 nm Cu, (c) 8 nm Cu, (f) 12 nm Cu and (g) 15 nm Cu thicknesses. (d) Graphical representation of the magnitude of the stopband blue-shift ($\Delta\lambda_{min}$) for Cu layer thicknesses over time. (e) EDX elemental analysis showing the relative wt% of elements detected in a newly coated Cu metal layer over a PS opal template as compared to elemental analysis of the same sample taken after 1 week of exposure to ambient atmosphere conditions.

The process of Cu metal layer oxidation and the optical effects associated with oxidation of the metal can be seen in Fig. 5. Figure 5 (a) schematically depicts the formation of a layer of copper oxide on the top surface of the metal layer after 1 week. We assume that some amount of the Cu metal is oxidised to various copper oxides after 1 week of exposure to an ambient atmosphere at room temperature. The accompanying SEM images of the surface of the metal-coated spheres appear to show an expanded and rougher film surface after this 1-week oxidation period. Elemental composition analysis is presented in a table in Fig. 5 (e), showing the relative wt% of elements detected from EDX line scan measurements for 20 nm of Cu metal deposited over the surface of polystyrene spheres. The elemental composition spectra and EDX line scan areas can be seen in more detail in the Supplementary Materials Fig. S11. The elemental composition is compared on the same sample surface between a recently deposited (~20 minutes after film deposition) measurement and a scan taken after allowing for 1 week of oxidation of the Cu film under ambient atmosphere conditions.

For the recently deposited Cu film, carbon and copper are present in highest abundance, as expected from copper metal sputtered over polystyrene $(C_8H_8)_n$. Oxygen is present in this initial scan, accounting for 2.5 wt% of the elements detected. Trace levels of silicon, chlorine and sulfur are also detected, most likely present due to the glass composition $(SiO_2)$ or residuals from the PS sphere manufacturing e.g. the sulfate ester which gives the PS spheres a slight negative charge. At least some of the initial oxygen detection can be attributed to various oxides in the glass or residual compounds on the PS spheres; slight oxidation of the metal film may also contribute to detected oxygen levels at this point. After 1 week of exposure to ambient conditions, the elemental composition of a similar area on the same sample shows almost equivalent levels of carbon, copper, silicon, chlorine and sulfur. The relative wt% of oxygen has almost doubled from 2.5% to 4.9%, suggesting a moderate oxidation of the copper film to various copper oxides, as expected.

The optical spectra of the Cu-modified PS opals, presented in Fig. 1, are examined following an oxidation period of 1 week in ambient atmosphere conditions. The spectra presented in Figs. 5 (b), (c), (f) and (g) are obtained and compared for measurements made on recently metal-coated samples versus samples allowed to oxidise for 1 week. Slight increases in transmittance are observed for most samples, to be expected from the partial removal of reflective metal particles. Interestingly, in the case of the (111) photonic stopband we observe a consistent reduction in the magnitude of the blue-shift caused by a metal coating. A bar chart summarising these results can be seen in Fig. 5 (d), showing a reduction in the blue-shift ($\Delta\lambda_{min}$) of the stopband for every metal layer thickness after a period of 1 week. An overall blue-shift of the stopband, compared to the uncoated opal stopband position, is still observed for every sample after oxidation. Copper-coated $TiO_2$ inverse opals were also examined after 1 week of oxidation, and detailed in the Supplementary Materials Fig. S8. The Cu-coated $TiO_2$ IO showed a dramatic reduction in the magnitude of the blue-shift from 36 nm to just 8 nm after 1 week of oxidation. It would appear that the specific properties of the metal (and not the metal oxide) are important for establishing the blue-shift of the (111) photonic stopband in a metallo-dielectric photonic crystal.

The placement of the metal material in the opal structure also seems to be significant for inducing the blue-shift in the (111) photonic stopband. Figures 6 (a) and (b) represent two methods in which metal material was placed in contact with PS opal photonic crystals. The scenario depicted in Fig. 6 (a) has been discussed for materials presented in Figs. 1, 3 and 5; metal layers are deposited over the top surface of the opal photonic crystal by sputter deposition. In addition to a conformal metal coating on the top surface of the opal (see Fig. 6(a)), we confirm that some metal deposited onto the second layer through the interstitial voids on the top surface. The presence of metal on the second layer of opals was confirmed via EDX line scan measurements of Ni and Au-coated PS opals, as seen in the Supplementary Materials Figs. S9 and S10. EDX measurements indicate the presence of reduced metal content, compared to the surface coating, in the interstitial voids atop the second layer of PS spheres. The EDX measurements for Ni also indicate a relatively uniform thickness for the Ni metal content deposited on the top surface.

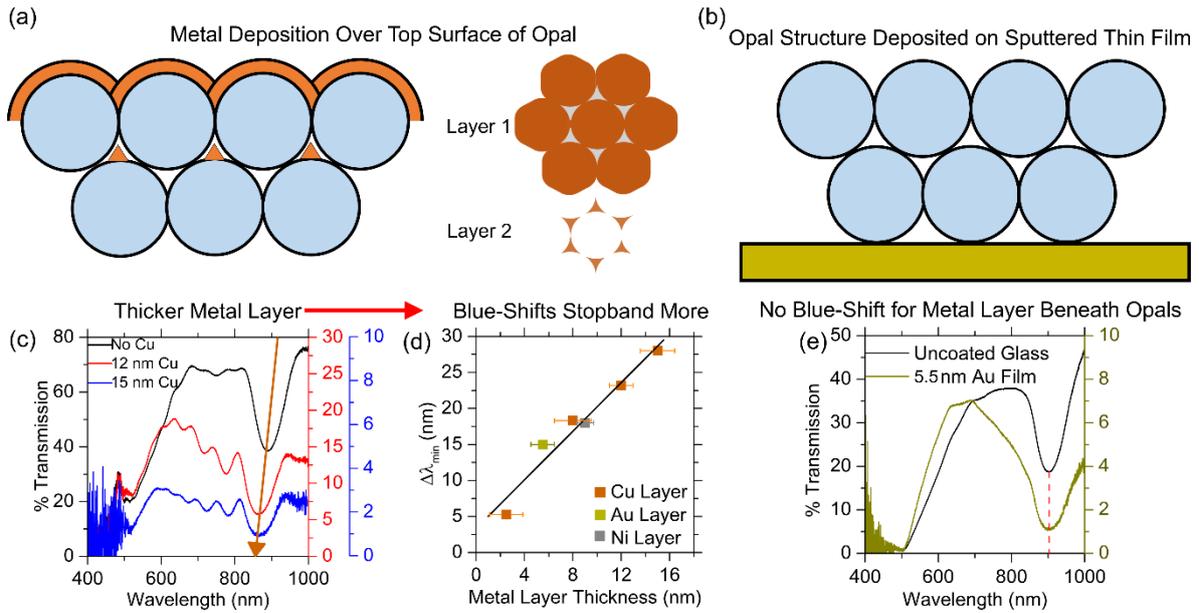

**Fig. 6** Schematic diagrams depicting the proposed position of metal in metallo-dielectric structures prepared from (a) sputter deposition over the top surface of the opal and (b) opal template formation over a sputtered metal film on a glass substrate. The blue-shift ($\Delta\lambda_{\min}$) in the stopband position for metal deposited over the surface of an opal is shown for transmission spectra and graphically in (c) and (d), respectively. (e) A transmission spectra comparison for opals prepared on an uncoated glass substrate versus a 5.5 nm Au metal film.

Figures 6 (c) and (d) summarise the observations made for PS opals coated with metal; a blue-shift of the (111) photonic stopband is observed for every metal investigated and the magnitude of the blue-shift increases with thicker metal layers, as seen with Cu metal. Figure 6 (b) illustrates the scenario of a PS opal structure grown on a sputtered metal film. The bottom of the photonic crystal structure is in direct contact with the metal; the metal is not incorporated into the structural voids or layers of the opal template in this case. Optical spectra showing an opal prepared on an uncoated substrate versus a gold film coated substrate, deposited with identical sputter coating settings to the gold film formed in Fig. 3 (c), can be seen in Fig. 6 (e). For consistency in the spectral analysis of the opal template, half of the substrate was coated with Au and the other half uncoated so that spectra could be taken from different areas of the same sample. Optically, a significant decrease in the transmission intensity for the Au-coated substrate is observed in Fig 6 (e), as expected due to the high reflectivity of Au metal. The (111) photonic stopband, located at 905 nm for the uncoated substrate, displays no detectable blue-shift when opals were deposited on the Au-coated substrate versus the uncoated substrate. This contrasts with the 15 nm blue-shift observed for Au-metal deposited over the top surface of the opal. Metallisation of the opal in the form of hemispheres, with some interstitial infilling, is necessary to invoke a stopband blue-shift; a planar gold film on one side of the opal photonic crystal slab has no effect, even when the metal is capable of supporting surface plasmon polaritons.

Regarding the blue-shift of the photonic stopband, we posit that the optical properties of the metal and its location in the dielectric medium are necessary considerations for causing a shift in the (111) stopband position. This blue-shift is detected for all metals (Cu, Ni and Au) tested and for both the artificial opal and inverse opal templates. We believe the most likely explanation for this behaviour is a reduction in the effective dielectric constant or effective refractive index of the metallo-dielectric photonic crystal. Previous works with various metallo-dielectric structures have also assumed a reduced effective dielectric constant or effective refractive index[67] [72] [74] [75]. The plasma frequencies for Cu, Ni and Au metal films are reported as energies in the UV range[83], with a negative real part of the dielectric function at energies below the plasma frequency. Metals at optical frequencies below the energy of the plasma frequency, would thus feature a negative dielectric constant. Incorporating a metal into a photonic crystal template should enhance the dielectric contrast of the material and reduce the effective dielectric constant of the composite material by replacing air in the structure with a negative dielectric material, and this occurs in the top surface region where the spheres are half-coated, and the interstices between spheres that are also metallized. In terms of the (111) photonic stopband, this reduction in the effective dielectric contrast of the photonic crystal structure is most likely reflected as a decrease in the effective refractive index of the structure, resulting in a blue-shift of the stopband position per Eq. (2). With this assumption, more metal introduced into the structure would act to further decrease the effective refractive index while simultaneously increasing the magnitude of the blue-shift. The thin surface coatings of metals applied in fabricating our metallo-dielectric structures would only replace a small fraction of the air in the material, establishing small changes in the effective refractive index. Taking the 12 nm Cu layer, 9 nm Ni layer and 5.5 nm Au layer deposited on PS opals (see Fig. 3) as an example, the stopband shifts of 23, 18 and 15 nm would constitute minor changes to the effective refractive index from 1.422 to 1.385, 1.393 and 1.398, respectively. Based on our investigation of data, it would seem as if thin coatings of a negative dielectric metal are sufficient to induce minor changes in the effective refractive index of photonic crystal structures, subsequently inducing a blue-shift in the (111) photonic stopband position.

It is important to consider the role of plasmons is these structures. Extraordinary optical transmission from coupling of surface plasmons will only be observable for these structures at higher wavelengths (beyond our measurement capability), if these modes are supported. The 3D photonic crystal (whether opal or inverse opal) defines the Bragg diffraction response and the stopband. With a hemispherical metallic coating on the top layer and partial, but periodic, infilling of the voids in the case of colloidal opals, coupling between a metallized 2D slab and the 3D photonic crystal is considered briefly. The higher energy oscillations observed here are also seen in metal coated 2D monolayer opal colloidal photonic crystals[84], and we also note a suppressed transmission caused by the larger imaginary component of the dielectric constant for Ni coated metallo-dielectric photonic crystals. However, observing these transmissions features in 3D opals is consistent with cavity modes. The metallization modifies the stopband in a unique manner (a blue-shift) even with an effective inverse in the periodicity of the dielectric constant in an inverse opal. Angle-resolved investigations using polarized light will likely be needed to determine any plasmonic coupling (Bragg and Mie plasmons) and a wider spectral range is necessary to confirm EOT effects or Fano resonances in metal-coated metallo-dielectric photonic crystals. The fact that we see this response for the first time in a metal-coated inverse opal, consistent with the mechanism that allows for these observations in colloidal opal photonic crystals, provides opportunities for investigations and applications in photocatalysis, spectroelectrochemical supports, and surface-enhanced scattering spectroscopy to name a few.

## Conclusions

The optical effects of metallo-dielectric photonic crystals have been investigated for a variety of different metals and structured dielectric materials. Simple metal surface coatings applied to both opal and IO templates were demonstrated to significantly alter the optical spectrum of the material in terms of the transmitted light intensity, the appearance of several diffraction resonances and a consistent shift in the position of the photonic stopband. Analysis of the various metallo-dielectric photonic crystals presented show that the thickness of the metal layer, the negative dielectric properties of the metal and the placement of the metal deposit in relation to the ordered structure are all essential considerations in predicting the optical changes to the modified system. Knowledge of these parameters should allow for a tuneable response to the optical spectrum of a metal-modified dielectric medium.

For Cu, Ni and Au metal deposited over the top surface of PS sphere opal templates, persistent blue-shifts in the photonic stopband position were recorded relative to uncoated PS sphere templates. Thicker metal coatings were correlated with increased magnitudes of the blue-shift in Cu metal. Expanding our analysis to investigate metal infiltration into innately different structured materials, the $TiO_2$ inverse opal, presented a corresponding blue-shift of the photonic stopband found in IO structures modified with Cu, Ni and Au. In $TiO_2$ IOs, stronger stopband shifts were observed for Cu and Au metal deposits compared to Ni, with possible hot electron generation altering the optical properties of the semiconductor. Nevertheless, a blue-shift of the photonic stopband position of some magnitude was observed for every metal deposited and incorporated into the structure of PS sphere opal and $TiO_2$ IO photonic crystals.

Through oxidation of the Cu metal, it was demonstrated that the specific properties of the metal material were necessary for the reported (111) stopband blue-shift. Oxidation of the Cu metal to various copper oxides decreased the reported magnitude of the blue-shift. Forming metallic hemispheres as the opal top coating is necessary to induce a stopband spectral blue-shift. By formation of the PS opal template on a Au-coated substrate and the subsequent detection of no stopband blue-shift, the placement of metal material directly into the photonic crystal material medium was linked to the origin of the spectral blue-shift. The consistent (111) photonic stopband blue-shift found using all metals used in this work is most plausibly explained by a reduction in the effective refractive index of the metallo-dielectric photonic crystal. This long-standing assumption has been presumed and attributed to observed spectral shifts in other works[67] [72] [74] [75]. With the results and evidence presented here, a strong argument can be made for clarification and association of the origin of the photonic stopband blue-shift in metallo-dielectric structures with a reduction in the effective refractive index of the composite structure. We posit that the inclusion of a negative dielectric metal in a photonic crystal structure is not only sufficient to increase dielectric contrast in the material, as was proposed in initial reports of metallo-dielectric structures[66] [67], but also to tune the effective dielectric constant or effective refractive index of the dielectric medium. Our results would indicate that even thin surface coats of the photonic crystal structure are capable of moderate modifications to the effective refractive index of the system, dictating the consistently observed stopband blue-shift with metal infiltration into the structure.

**Acknowledgments**

We acknowledge support from the Irish Research Council Government of Ireland Postgraduate Scholarship under award no. GOIPG/2016/946. This publication has also emanated from research supported in part by a research grant from SFI under Grant Number 14/IA/2581. We also acknowledge funding from the Irish Research Council Advanced Laureate Award under grant no. IRCLA/2019/118.

**Conflict of interest**

The authors declare no conflict of interest.

**Author information**

Corresponding author: Colm O'Dwyer, email: c.odwyer@ucc.ie; Tel: +353 21 490 2732

**Supporting Information for**

# Metallo-Dielectric Photonic Crystals and Bandgap Blue-Shift

Alex Lonergan[1], Breda Murphy[1], and Colm O'Dwyer[1,2,3,4]*

*[1]School of Chemistry, University College Cork, Cork, T12 YN60, Ireland*

*[2]Micro-Nano Systems Centre, Tyndall National Institute, Lee Maltings, Cork, T12 R5CP, Ireland*

*[3]AMBER@CRANN, Trinity College Dublin, Dublin 2, Ireland*

*[4]Environmental Research Institute, University College Cork, Lee Road, Cork T23 XE10, Ireland*

Size distributions for opal diameters used in the deposition of Cu metal layers onto the surface of artificial polystyrene opal templates can be seen in Fig. S1. For metallo-dielectric opals, the mean diameters for $D_{sphere}$ and $D_{shell}$, as defined in the main text, are estimated for each template from a large selection of SEM images of the metal layers on the opal surface. The thickness of the metal layer is defined as half of the difference between $D_{shell}$ and $D_{sphere}$, as shown in Fig. 1 in the main text. For the size distributions shown in Fig. S1, longer sputter deposition times were used to successively deposit thicker layers of Cu metal on the surface of the opal. As measured from the mean diameters of $D_{sphere}$ and $D_{shell}$, the metal layer thickness calculated for each of the 4 Cu metal layers deposited are (a) 2.5 nm, (b) 8 nm, (c) 12 nm and (d) 15 nm. Small SEM images are inset in each of the size distributions, showing the measurement definition for $D_{sphere}$ (blue arrow) and $D_{shell}$ (orange arrow) for each template. The small standard deviations (SD) on each of the size distributions are indicative of the monodispersed polystyrene particles and uniform metal coat thicknesses deposited on the structures.

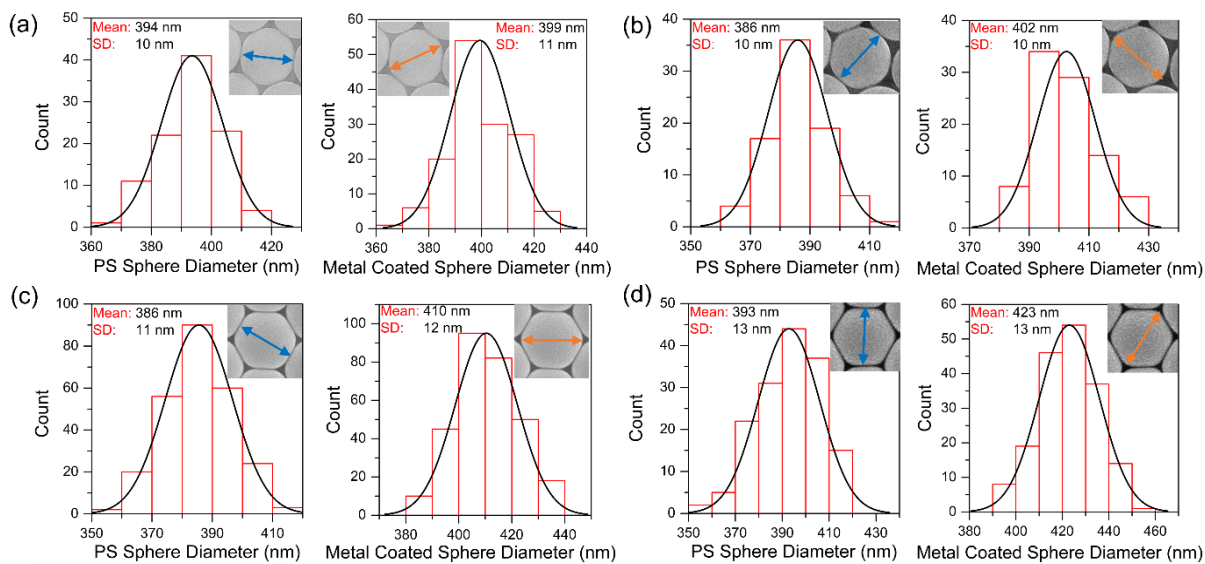

**Fig. S1** Size distributions with measurements from SEM images of hexagonal metal shells calculating mean diameters for the uncoated opal ($D_{sphere}$, side-to-side) versus Cu metal coated opals ($D_{shell}$, vertex-to-vertex). Measurements show thicker metal layers for longer metal sputter deposition times with (a) 2.5 nm, (b) 8 nm, (c) 12 nm and (d) 15 nm Cu metal layers calculated.

SEM images showing the appearance of the top surface of Cu metal layers deposited on polystyrene opal templates are included in Fig. S2. Images of these Cu metallo-dielectric structures are shown for calculated metal layer thicknesses of (a) 2.5 nm, (b) 8 nm, (c) 12 nm and (d) 15 nm. From a 2D SEM image, the hexagonal shape of the metal coating, as proposed in Fig. 1 in the main text, becomes increasing more pronounced with thicker metal layers deposited on the surface. The expansion of the growing metal shell into the interstitial air region reduces the observed interstitial area and lends the appearance of hexagonal shape observed from SEM imagery. From the SEM images of the slightly modified 2.5 nm Cu layer in Fig. S2 (a), the structure still appears spherical with large interstitial air regions still visible. In contrast, the SEM images of the thicker 15 nm Cu layer in Fig. S2 (d) show the hexagonal geometry of the top surface and heavily reduced interstitial air regions.

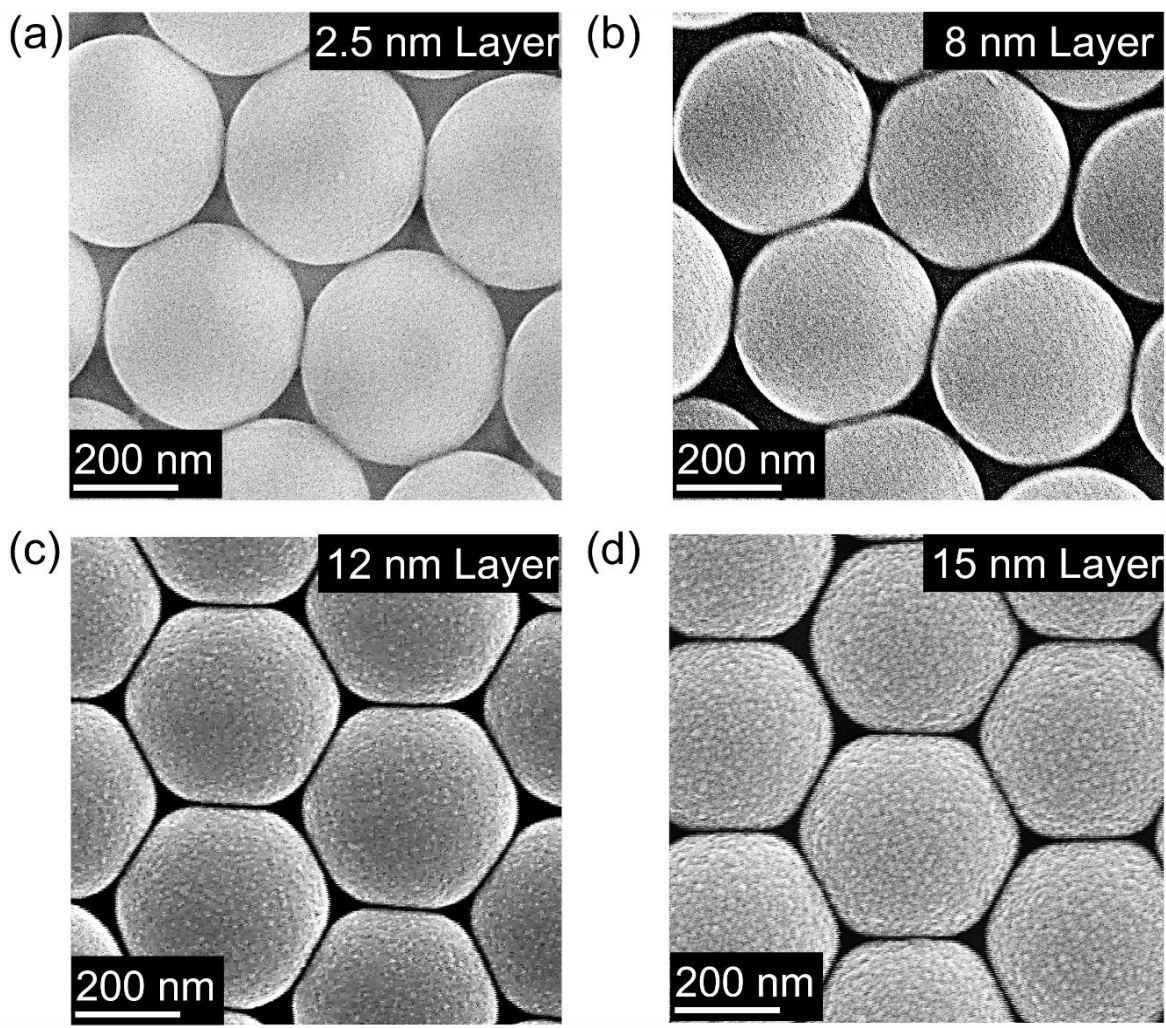

**Fig. S2** SEM images of the top surface of opals covered with Cu metal layers of calculated thicknesses (a) 2.5 nm, (b) 8 nm, (c) 12 nm and (d) 15 nm.

A two-dimensional SEM characterisation of the alterations to the surface geometry of metallo-dielectric opals is detailed in Fig. S3. Specifically, the growth of the metal shell with increasing metal layer thickness is linked to the reduction of the observed interstitial air region in SEM images. A qualitative example from SEM images can be seen in Fig. S3 (a), where the deposition of a thicker Cu metal shell visually reduces the interstitial area observed. A clearer depiction of the changes to the interstitial air region with increasing metal layer thickness is shown in Fig. S3 (b) using SEM images with increased brightness and contrast. The dark spaces represent the interstitial areas observed from SEM images for a bare opal template, a 12 and 15 nm Cu metal layer thickness. The hexagonal shape of the top surface becomes much more visible with increasing metal layer thickness. The interstitial areas are drastically reduced with increasing metal content; calculated estimates predict interstitial areas of 31% and 22%, compared to the uncoated opal template, for the 12 and 15 nm Cu layers, respectively. A schematic representation of the reduction of the interstitial region with increasing metal shell thickness is depicted in Fig. S3 (c). In this schematic, the interstitial region (shown in white) is reduced from its initial size (shown in blue) due to the presence of a metal layer (orange material) over the surface of the opal.

For an assumed equal distribution of metal across a growing metal shell on the surface of an opal, the geometric reduction of the 2D interstitial area can be modelled and calculated based on the area of intersection of three tangent circles. Uncoated opals would be perfectly tangential and represent 100% of the available interstitial area, $A_i$. The ratio of the increased diameter (metal shell, $D_{shell}$) versus the initial diameter (opal diameter, $D_{sphere}$) can be used to quantify the reduction in interstitial area. Using some geometry, the general solution of for the reduced 2D interstitial area, relative to the initial interstitial area, created by a growing metal shell can be found with knowledge of the using:

$$A_i = \frac{\left[ 2\sin 60 - \frac{3}{2}\left(\frac{D_{shell}^2}{D_{sphere}^2}\right) \left(\left(\frac{60 - 2\cos^{-1}\left(\frac{D_{sphere}}{D_{shell}}\right)}{180}\right)\pi - 3\left(\frac{D_{shell}}{D_{sphere}}\right)\sin\left(\cos^{-1}\left(\frac{D_{sphere}}{D_{shell}}\right)\right)\right) \right]}{\left[ 2\sin 60 - \frac{\pi}{2}\right]} \ldots (1)$$

Using Eq. (1), a plot of the reduced interstitial area is plotted against the ratio of the metal shell to the initial opal diameter in Fig. S3 (d). The reduced interstitial area for several Cu metal shells produced are marked for reference on this curve. Using this method, simple SEM measurements of the diameter of structures can be used to characterise the 2D surface appearance of the metallo-dielectric structure.

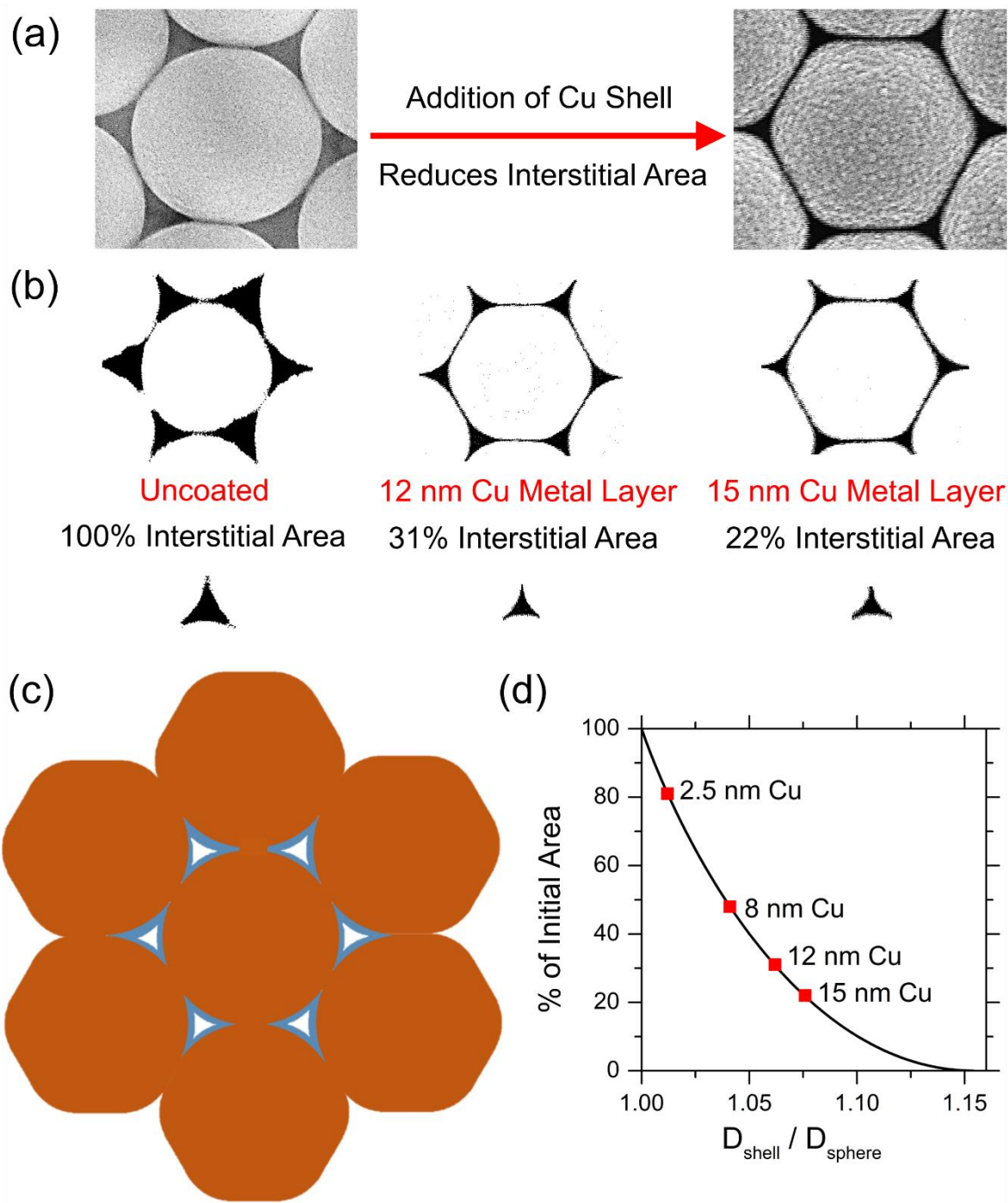

**Fig. S3** (a) SEM images showing the observed reduction to the interstitial on the surface of opal structures with a metal layer deposited on the top surface. (b) SEM images with image brightness and contrast adjusted to depict the sizes of the interstitial for a bare opal, an opal with a 12 nm Cu layer and an opal with a 15 nm Cu layer. (c) Schematic representation of the reduction in the interstitial area with uniformly growing metal layers on an opal surface. (d) Calculated reduced interstitial area, as per Eq. (1), for a thickening metal shell ($D_{shell}$) on a spherical opal ($D_{sphere}$). Interstitial areas are highlighted for Cu layers measured from SEM analysis.

The structure of the Ni metal coated polystyrene opal template is characterised in Fig. S4. The SEM images in Figs. S4 (a) and (b) show the respective appearances of the opal surface before and after the deposition of the Ni metal. Much like in the case of the Cu metal, the deposition of Ni metal is seen to decrease to the interstitial air area of the structure and create the 2D hexagonal appearance associated with the metal shell growth. Size distributions for (c) $D_{sphere}$ and (d) $D_{shell}$ are shown for the Ni metallo-dielectric opal. The metal layer thickness calculated for this particular structure indicates a Ni metal layer of 9 nm deposited on the surface of the opal.

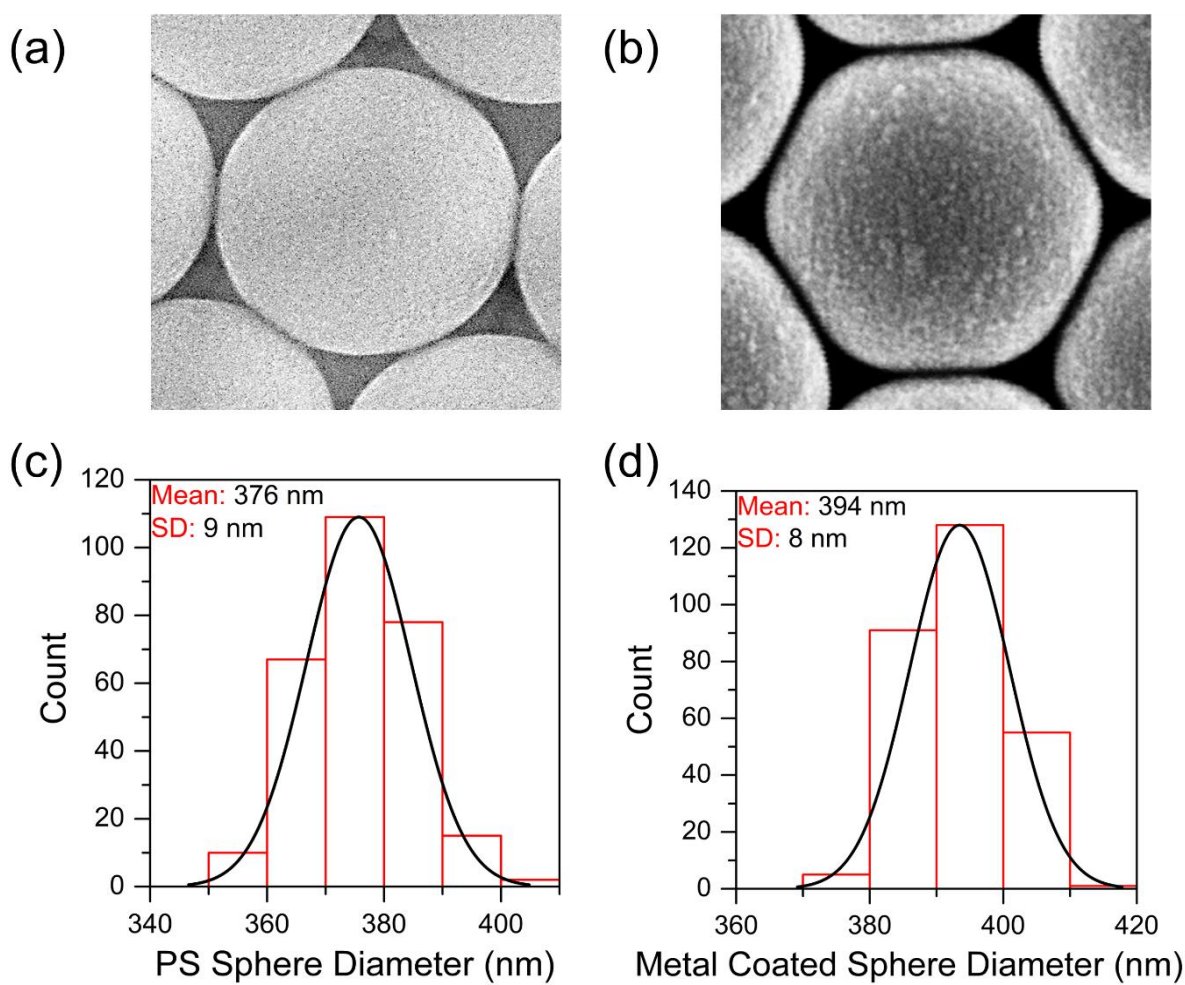

**Fig. S4** SEM images depicting the surface of (a) an uncoated opal and (b) an opal coated with a 9 nm thick nickel metal layer. Size distributions calculating the mean diameter for (c) $D_{sphere}$ and (d) $D_{shell}$ for the template used for nickel metal deposition.

The structure of the Au metal coated polystyrene opal template is characterised in Fig. S5. The SEM images in Figs. S5 (a) and (b) show the respective appearances of the opal surface before and after the deposition of the Au metal. Much like in the previous cases of Cu and Ni metal, the deposition of Au metal is seen to decrease to the interstitial air area of the structure and create the 2D hexagonal appearance associated with the metal shell growth. The individual particle sizes for the Au layer appear to be larger than the Cu and Ni particles, as observed from the relevant SEM images. Size distributions for (c) $D_{sphere}$ and (d) $D_{shell}$ are shown for the Au metallo-dielectric opal. The metal layer thickness calculated for this Au-coated opal is found as 5.5 nm.

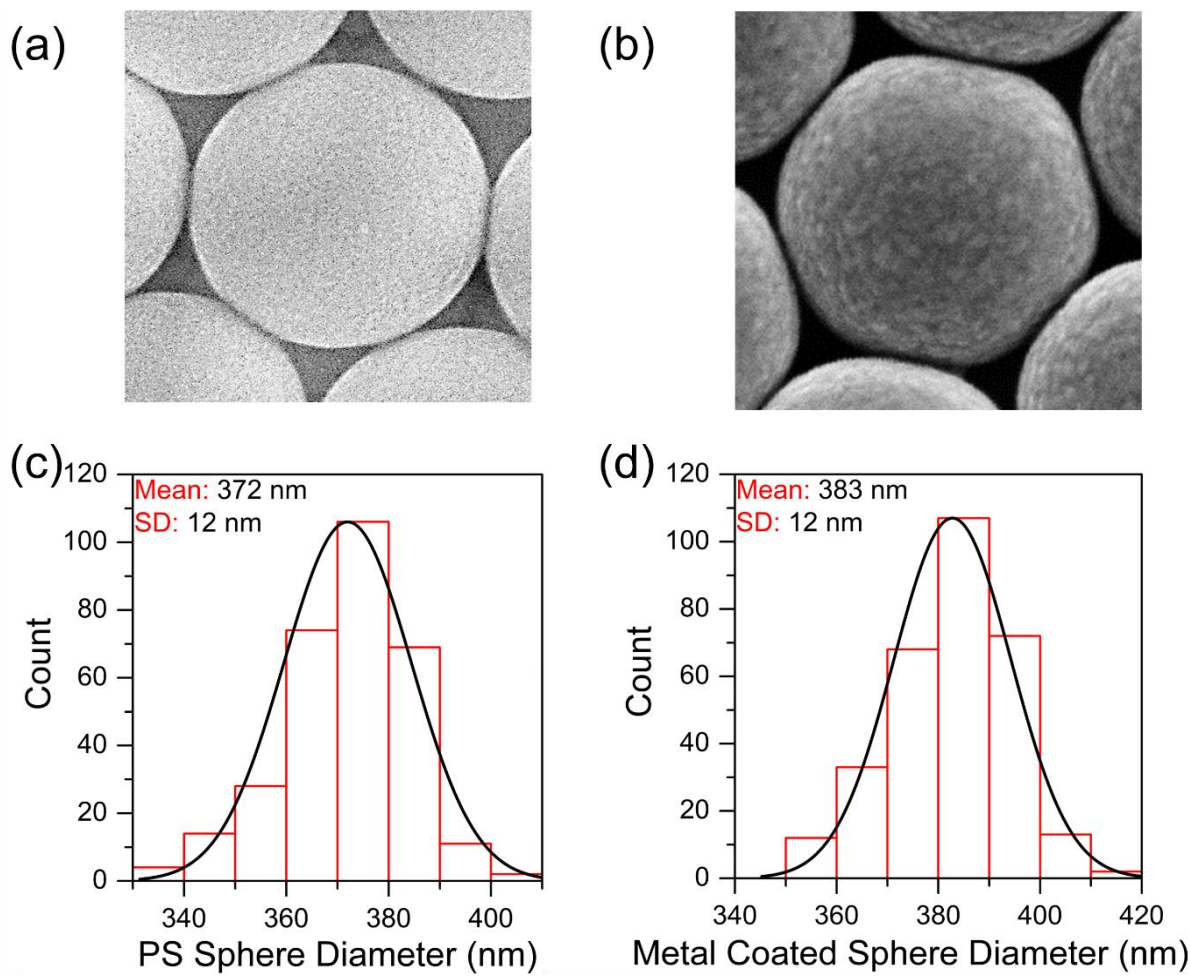

**Fig. S5** SEM images depicting the surface of (a) an uncoated opal and (b) an opal coated with a 5.5 nm thick gold metal layer. Size distributions calculating the mean diameter for (c) $D_{sphere}$ and (d) $D_{shell}$ for the template used for gold metal deposition.

The characterisation of a typical TiO₂ inverse opal used in this study is detailed in Fig. S6. The planar (111) geometry of the top surface of the IO is shown in Fig. S6 (a). Layers beneath the top surface are also visible due to highly porous structure associated and expected from the IO. TiO₂ IOs were prepared via sol-gel infiltration of a polystyrene opal of approximately 500 nm diameter. Calcination removed the opal template and crystallised the TiO₂ in the interstitial locations. Just like in the case of polystyrene opals, many SEM images were examined to calculate a mean centre-to-centre distance (D) which could be applied to optical estimations of the photonic stopband position. The size distribution of the mean centre-to-centre pore distance can be seen in Fig. S6 (b), showing a mean value of 440 nm. The standard deviation of 23 nm for the TiO₂ inverse opal indicates a slightly higher variation in feature size associated with the IO structure compared to the opal templates examined previously.

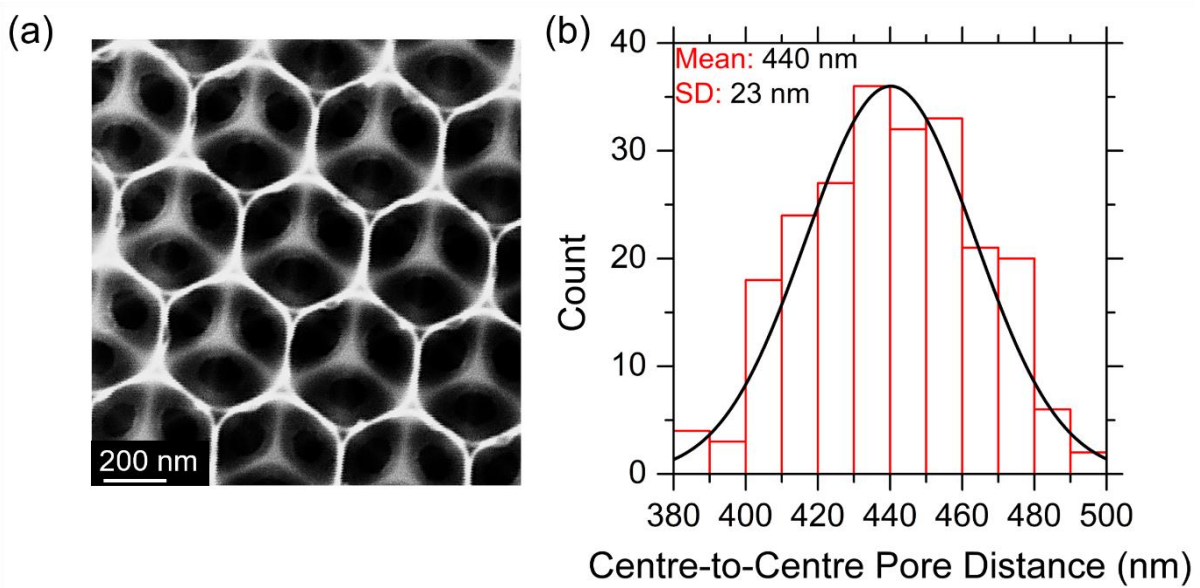

**Fig. S6** (a) SEM image of the surface of a typical TiO₂ inverse opal with subsequent layers of material visible underneath the top layer. (b) Size distribution analysis of the centre-to-centre pore distance for a typical TiO₂ IO, showing a mean calculated value.

Further structural characterisation of the TiO₂ IOs used in Ni and Au metal deposition can be found in Fig. S7. Production of the IO structure through sol-gel methods often leads to the formation of cracks in the IO film; individual "islands" on IO material form on the surface of the substrate. The typical appearances and sizes of these IO islands and cracks in the structure for the Ni and Au-coated $TiO_2$ templates are seen depicted in Figs. S7 (a) and (b), respectively. From these images, islands dimensions of $180 \times 95 \ \mu m^2$ and $110 \times 40 \ \mu m^2$ can be seen for the $TiO_2$ IO templates used for the Ni and Au metal deposition, respectively. The number of layers of IO material, constituting the height of the IO film, can be observed from regions on the edge of these islands.

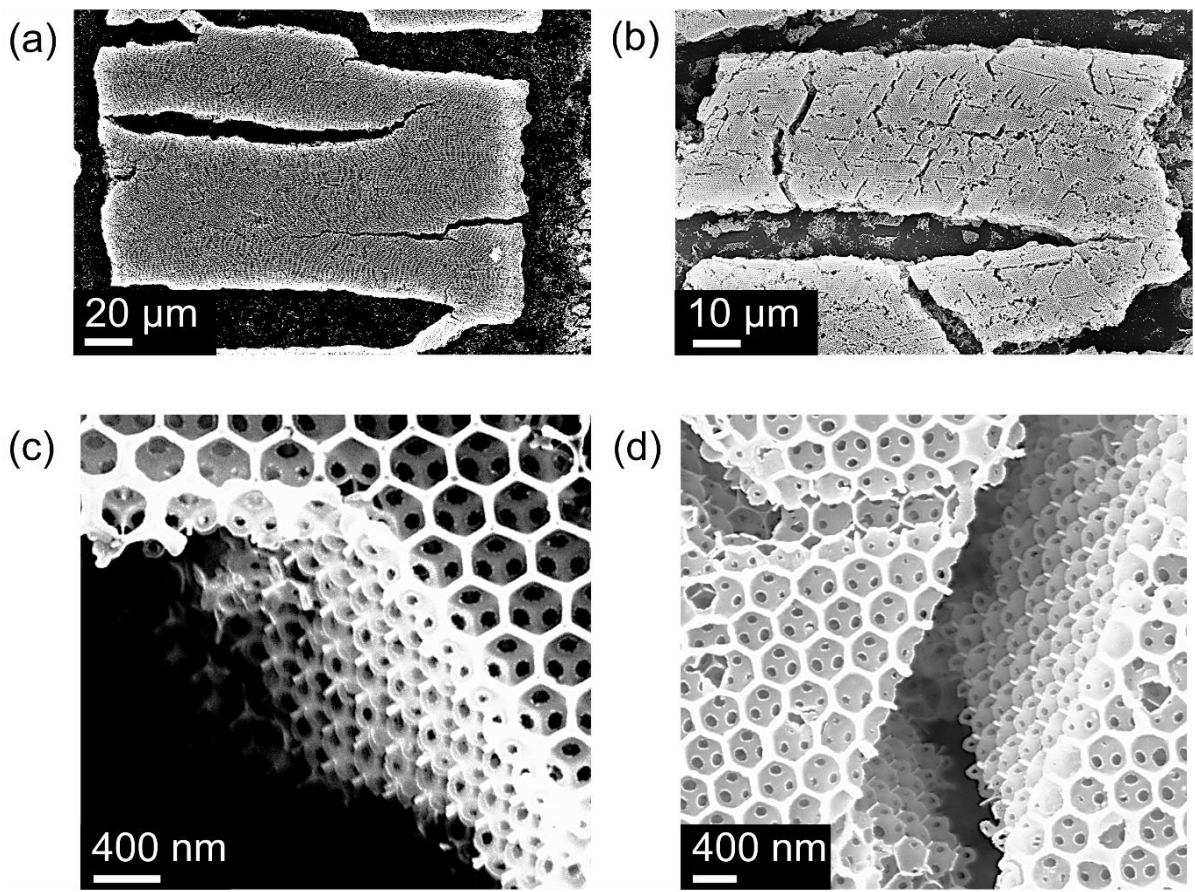

**Fig. S7** SEM images depicting the size of the individual "islands" of IO material comprising the $TiO_2$ inverse opal structure used for (a) Ni and (b) Au metal deposition. SEM images showing the number of layers of $TiO_2$ IO material, as seen from a crack in an IO "island", for IOs used for (c) Ni (~ 8 layers) and (d) Au (~ 11 layers) metal deposition.

Figs. S7 (c) and (d) display SEM images, with the number of layers of IO material visible, for the respective Ni and Au-coated $TiO_2$ IOs. Approximately 8 layers of IO material can be seen for the Ni-coated $TiO_2$ IO; 11 layers are visible for the Au-coated $TiO_2$ IO. The structural dimensions reported here for the $TiO_2$ IOs are largely typical of $TiO_2$ IOs prepared with this method. Similar dimensions were observed for the $TiO_2$ IOs used for Cu metal deposition (not shown).

The effects of Cu metal oxidation on the magnitude of the photonic stopband blue-shift on a $TiO_2$ IO template is assessed in Fig. S8. An SEM image of the surface of the IO covered with Cu metal is shown in Fig. S8 (a). The Cu metal appears to be distributed relatively evenly over the IO surface with similar features observed for each of the pores visible on the SEM image. In comparison with the bare $TiO_2$ IO image in Fig. S6 (a), the Cu-coated IO shows thicker walls and metal can be seen to penetrate further into the structure. Less of the layers underneath the top layer in the Cu-coated IO are visible, when compared to the uncoated IO structure, it's likely that the thicker walls of the Cu-modified structure cover more of the structure in the image. A comparison of the transmission spectra for recently coated Cu metal on a $TiO_2$ IO versus that same $TiO_2$ after 1 week in an ambient atmosphere environment is shown in Fig. S8 (b).

The features observed from these optical spectra match the behaviour reported for Cu-coated opals in Fig. 5 of the main text. After 1 week under ambient atmosphere conditions, some oxidation of the Cu metal on the surface of the photonic crystal is likely to occur. This oxidation of the metal layer slightly increases the transmission intensity of the sample while also appearing to further modify the position of the photonic stopband. For the $TiO_2$ IO, the change in the photonic stopband position is significant, moving from 651 nm to 679 nm after 1 week of oxidation under ambient conditions. The initial stopband position for the uncoated $TiO_2$ IO was located at 687 nm; the stopband position of both spectra shown in Fig. S8 (b) are blue-shift with respect to the uncoated IO stopband position. The magnitude of the blue-shift is dramatically decreased after 1 week of oxidation for the Cu metal on the surface, decreasing from a 36 nm shift to just an 8 nm shift. Reductions to the stopband blue-shift were also reported for Cu-coated opals in the main text; however, the magnitude of this decrease appears more significant for the TiO$_2$ IO structure. We attribute this effect to the wider surface area available for metal deposition with the IO structure. The increased surface deposition should allow a higher quantity of metal to be exposed and oxidised under ambient conditions.

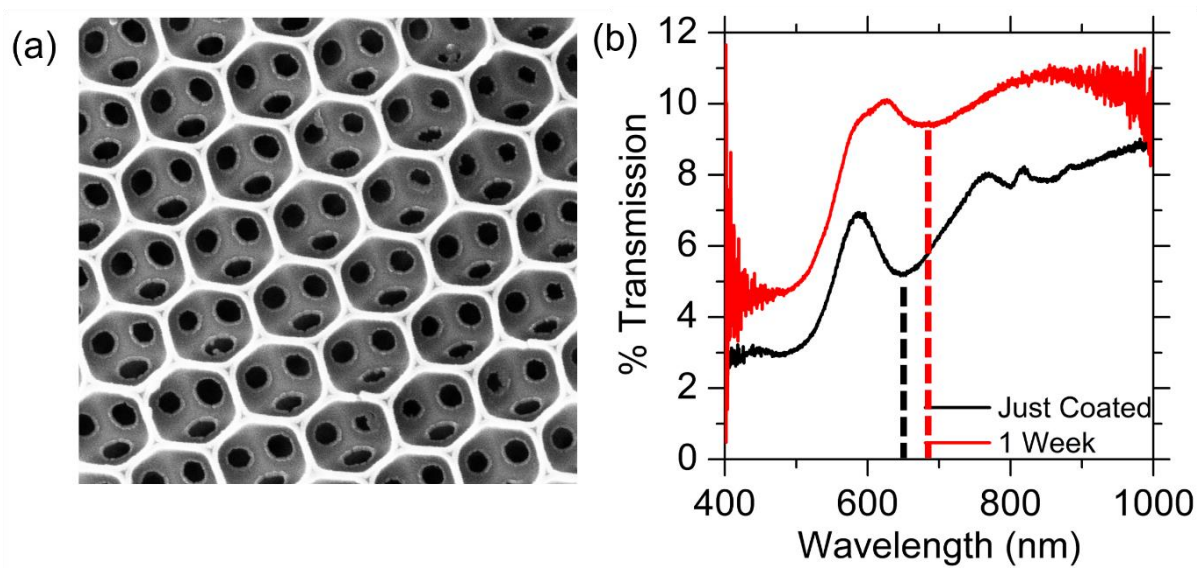

**Fig. S8** (a) SEM image of a TiO$_2$ IO surface with Cu metal deposited over the top of the material. (b) Transmission spectra comparison between the optical spectrum TiO$_2$ IO with a recently coated Cu metal layers versus the optical spectrum of the same TiO$_2$ IO taken after 1 week under ambient atmosphere conditions. The significant shift in the (111) photonic stopband position is highlighted.

The consistency of the metal coatings on the surface of an opal could was probed using energy dispersive X-ray spectroscopy (EDX) line scan measurements. Tracking the relative intensity counts of specific elements across the surface of the structure can provide information on the distribution of a particular element. The element for a metal coating was tracked and compared to the carbon content (effectively polystyrene, $(C_8H_8)_n$) to monitor the consistency of the metal coating versus the polystyrene template. Two EDX line scan measurements taken from the same 9 nm Ni-coated opal sample are shown in Figs. S9 (a) and (b). The line scan measurement in Fig. S9 (a) is directed across several interstitial air regions visible on the surface of the material.

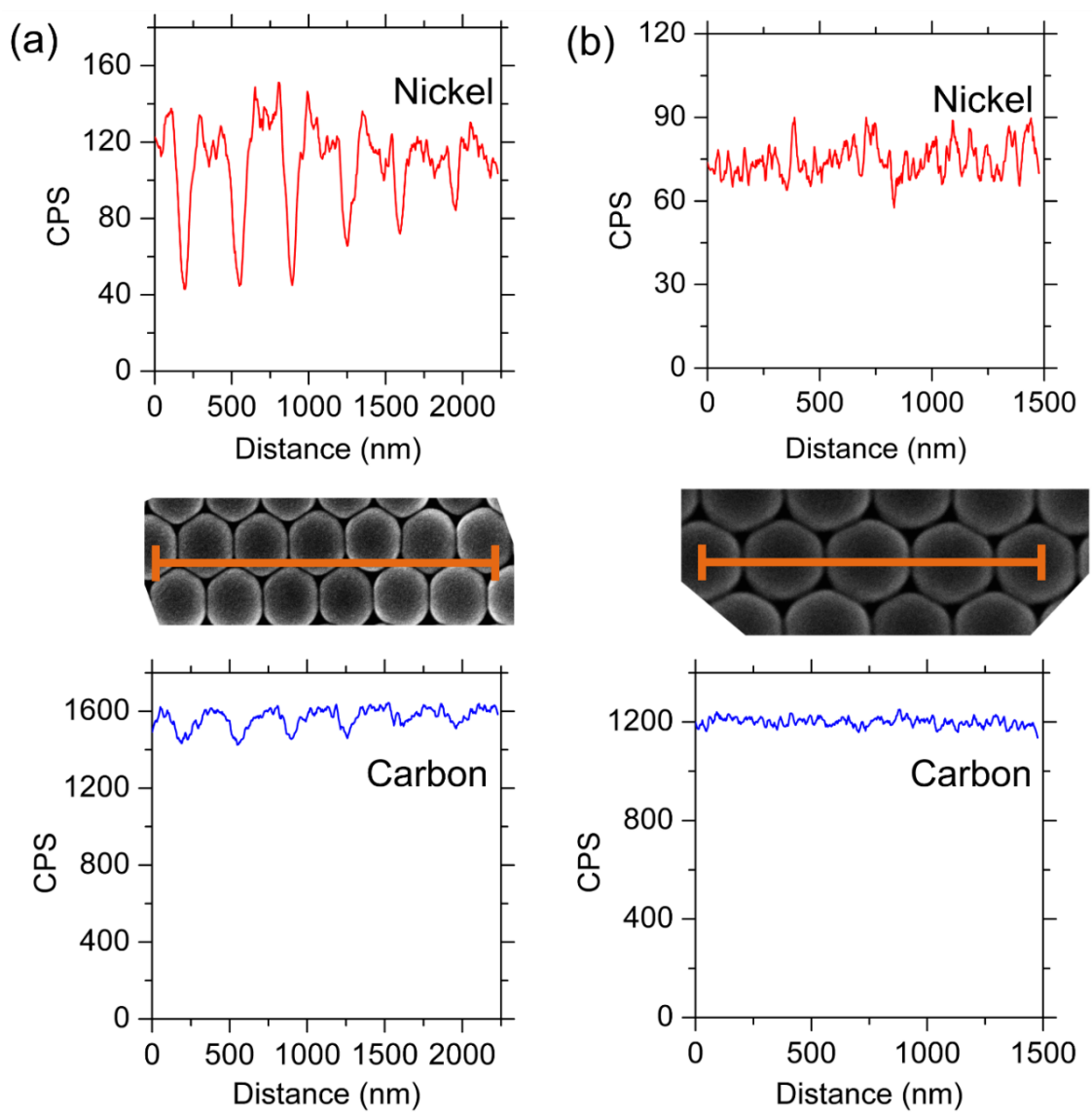

**Fig. S9** Nickel and carbon content detected via EDX line scan measurements across the surface of a 9 nm thick Ni metal layer deposited on an opal surface, alongside an SEM image of the line scan, for (a) a line scan over the interstitial area region of the opal and (b) a line scan over the touching centres of adjacent opals.

The nickel and carbon content plots both experience drops in intensity coinciding to the positions of these interstitial regions on the surface. The intensity drops are less pronounced for carbon, presumably due to the multilayer polystyrene opal structure. The nickel content exhibits significant drops in intensity at the interstitial points; however, the nickel intensity is non-zero valued even in the interstitial regions. This data would suggest that some metal content is indeed deposited on the surface of the second layer of opals in the structure, as posited in Fig. 6 (a) of the main text. A line scan measurement over the centre surface of adjacent opals is shown in Fig. S9 (b). In this case, the intensity for both nickel and carbon elements is relatively constant across the surface. This data would suggest that a roughly uniform thickness of metal comprises the metal layer deposited on the surface of the opal, as was assumed in the main text.

The Au content deposited on an opal surface is also investigated for a 5.5 nm Au metal layer deposited on a polystyrene opal using EDX line scan measurements, as seen in Fig. S10. In this instance the line scan is performed over a region with a defect vacancy present on the top layer of the opal template. This type of measurement provides an insight into the deposition depth of the Au metal into the opal template through a large vacancy in the surface layer. For carbon and gold elements, the lowest intensity region is located in the centre of the vacancy with overlap of the interstitial areas of lower layers. The content of gold varies significantly over vacancy region. A notable reduction in the gold content is detected on the second layer of opals visible in the SEM image. In spite of the surface opal vacancy, it would appear that only a limited amount of Au metal reaches the second layer of opals in the template. In fact, the amount of Au metal detected on the uncovered opals of the second layer (400 – 700 nm) is comparable to the amount of Au metal deposited on the second layer of opals underneath a standard interstitial air region (~ 1050 nm). From EDX analysis, it would seem that metals deposited on opal templates preferentially form metal shells on the surfaces of the opal with minor metal deposits on the second layer of the structure.

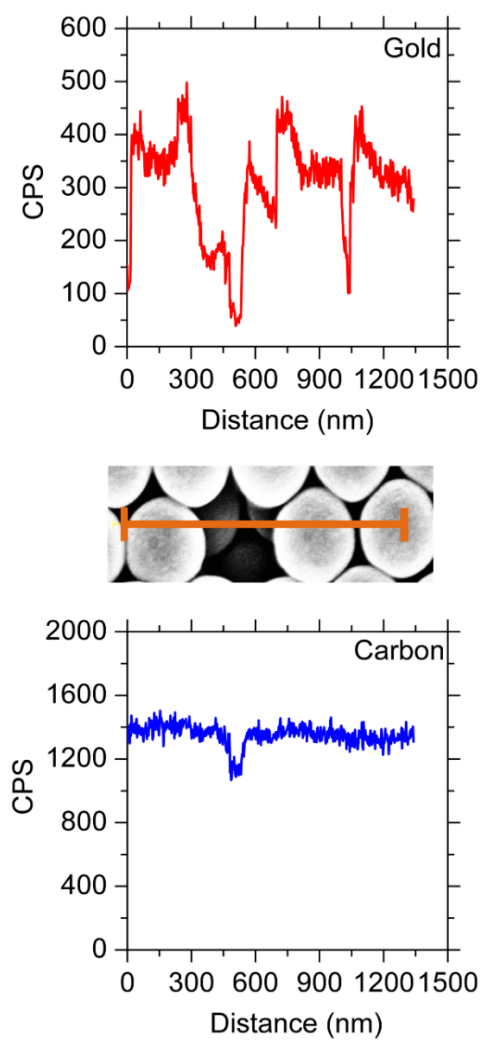

**Fig. S10** Gold and carbon content detected via an EDX line scan measurement across several opals and gap across the top surface for a 5.5 nm Au metal layer deposited on opal surface. The SEM image of the line scan region is also shown.

To demonstrate the increased oxygen content, attributed to copper oxide formation upon prolonged exposure to ambient atmosphere conditions, the copper metal film deposited over the surface of PS opals was analysed using EDX analysis. EDX elemental analysis was performed on a recently coated sample and compared to analysis from the same sample after a 1 week of exposure to ambient atmosphere. The elemental composition detected from two line scan datasets were presented in the main text Fig. 5 (e). Figures S11 (a) and (b) display the line scan images and the elemental composition spectra for analysis taken from a recently coated Cu film and a Cu films allowed to oxidise for 1 week, respectively. As commented in the main text, the relative wt% of oxygen detected has increased from 2.5% to 4.9% after 1 week; EDX elemental analysis would appear to indicate a higher oxygen content after exposure to ambient atmosphere, supporting the idea that copper oxides are forming on the surface of the film.

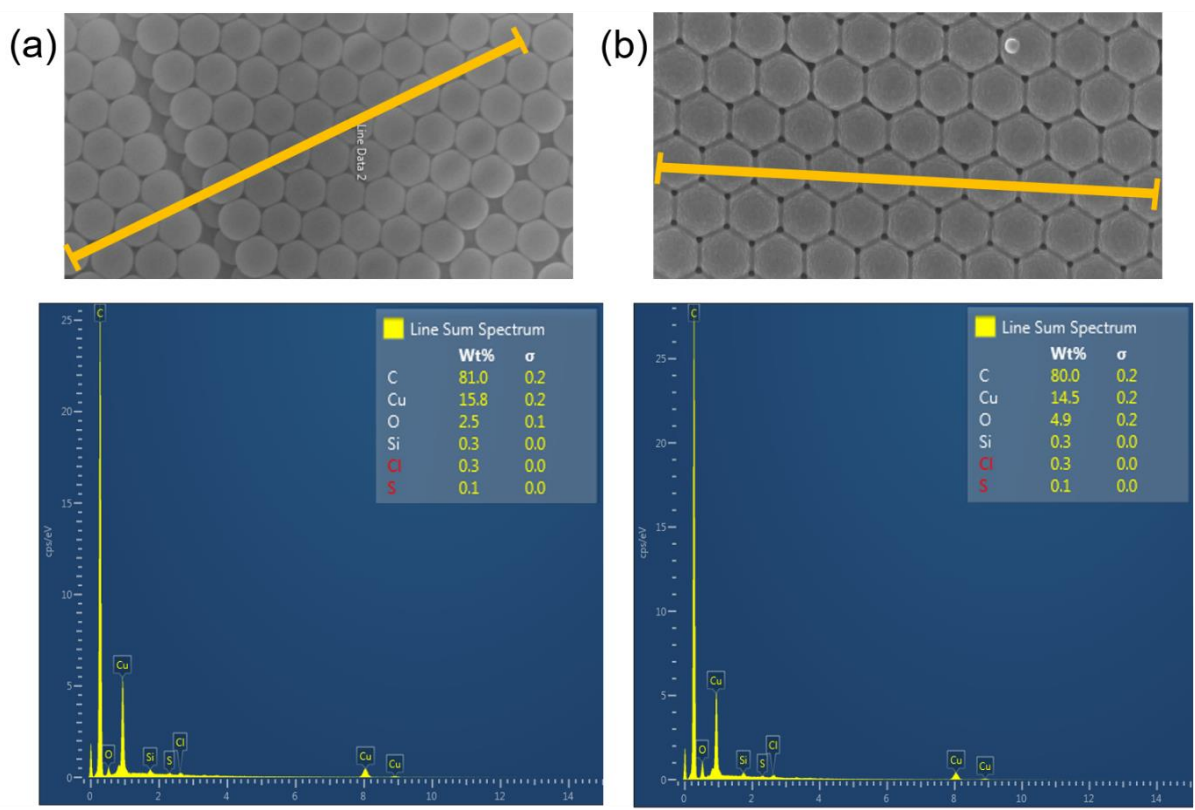

**Fig. S11** EDX line scan images and elemental composition spectra for 20 nm of Cu deposited on the surface of PS spheres for (a) a recently deposited Cu film and (b) the same Cu film after an oxidation period of 1 week.